\documentclass[table]{article}

    \usepackage[preprint]{neurips_2025}

\usepackage[utf8]{inputenc} 
\usepackage[T1]{fontenc}    
\usepackage{hyperref}       
\usepackage{url}            
\usepackage{booktabs}       
\usepackage{amsfonts}       
\usepackage{nicefrac}       
\usepackage{microtype}      
\usepackage{graphicx}
\usepackage{tcolorbox}
\usepackage{amsmath}
\usepackage{multirow}
\usepackage{xcolor}
\definecolor{lightgray}{gray}{0.9}
\definecolor{mygray}{gray}{0.92}
\newcommand{\cmark}{\ding{51}} 
\newcommand{\xmark}{\ding{55}}  
\definecolor{baselinecolor}{gray}{0.8}
\usepackage{tabularx}
\usepackage{multicol}
\usepackage{listings}
\usepackage{pifont} 
\usepackage{wrapfig}     
\usepackage{colortbl}

\definecolor{lightgray}{gray}{0.9}

\makeatletter

\newcommand{\Rmnum}[1]{\expandafter\@slowromancap\romannumeral #1@}
\makeatother

\newcolumntype{x}[1]{>{\centering\arraybackslash}p{#1pt}}
\newcolumntype{y}[1]{>{\raggedright\arraybackslash}p{#1pt}}
\newcolumntype{z}[1]{>{\raggedleft\arraybackslash}p{#1pt}}

\usepackage{subcaption}
\usepackage{wrapfig}
\usepackage[accsupp]{axessibility}
\usepackage{longtable}
\usepackage{fancybox}
\usepackage{geometry}
\title{MedTVT-R1: A Multimodal LLM Empowering Medical Reasoning and Diagnosis}

\author{Yuting Zhang, Kaishen Yuan, Hao Lu, Yutao Yue, Jintai Chen, Kaishun Wu$^{\ddagger}$\\
The Hong Kong University of Science \& Technology (Guangzhou) \\
yzhang430@connect.hkust-gz.edu.cn, wuks@hkust-gz.edu.cn\\
$^\ddagger$ Corresponding author
}

\begin{document}

\maketitle

\begin{figure}[h]
\centering
\includegraphics[width=1\linewidth]{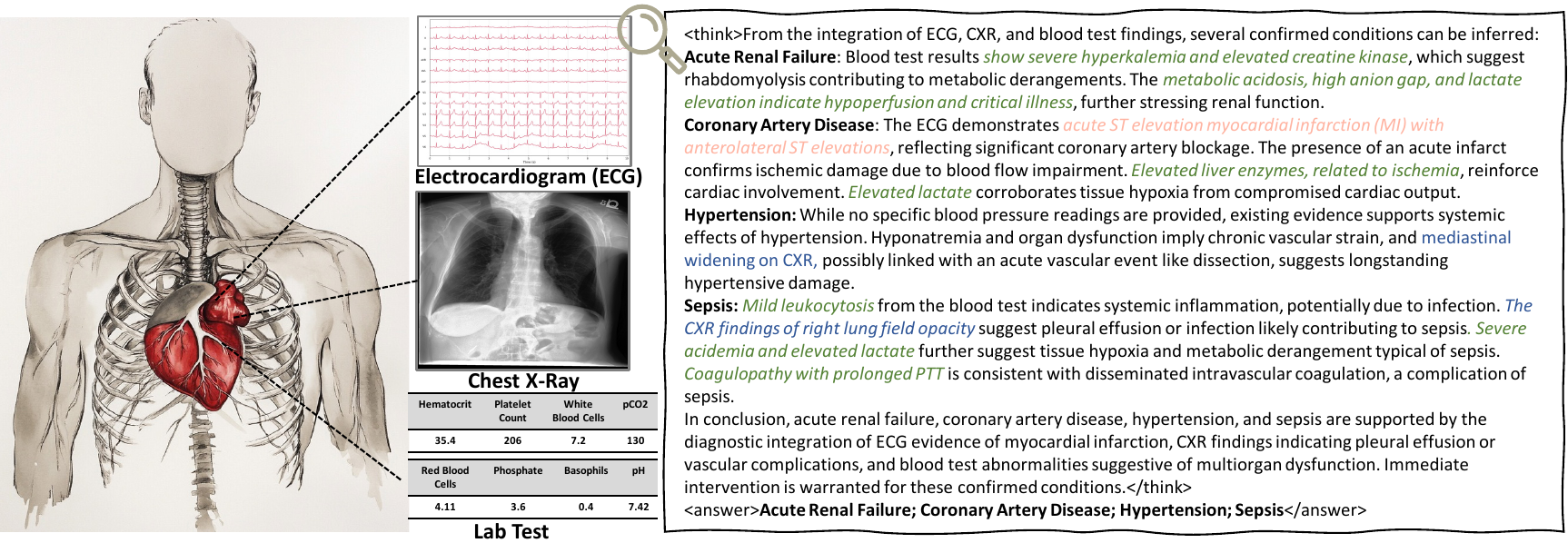}
\vspace{-1em}
  \caption{\small{Overview of \textbf{MedTVT-R1}: MedTVT-R1 seamlessly integrates Electrocardiogram (\textbf{T}ime Series), Chest X-ray (\textbf{V}isual Image), and Blood Test (\textbf{T}abular Data) to deliver comprehensive long-text \textbf{Med}ical reasoning and diagnosis across various diseases. }}
\label{fig:top_fig}
\end{figure} 

\begin{abstract}
 Accurate and interpretable multi-disease diagnosis remains a critical challenge in medical research, particularly when leveraging heterogeneous multimodal medical data. Current approaches often rely on single-modal data, limiting their ability to comprehensively understand complex diseases. To address this, we propose MedTVT-R1, a novel Multimodal Large Language Model (MLLM) framework designed to integrate clinical multimodal data for reasoning and diagnosing multiple diseases. We construct MedTVT-QA, a curated instruction dataset that provides question-answer pairs for physiological-level interpretations and disease-level diagnoses with a Chain of Evidence approach. MedTVT-R1 incorporates a modality perception layer to capture inter-modal dependencies and adaptively weight modality contributions. Additionally, we employ Group Relative Policy Optimization (GRPO)-based Reinforcement Fine-Tuning with a Jaccard Reward function to enhance diagnostic reasoning. Experimental results demonstrate MedTVT-R1's superiority in multimodal feature utilization and multi-disease diagnosis, offering significant potential for clinical applications such as diagnostic report generation and comorbidity reasoning. The dataset and code are available at \url{https://github.com/keke-nice/MedTVT-R1}.
\end{abstract}

\section{Introduction}
\label{sec:intro}

The rapid development of artificial intelligence (AI) has profoundly reshaped the landscape of medical research and clinical practice, especially in demonstrating significant progress and potential in medical data analysis~\cite{ccalli2021deep, liu2021deep, hernandez2022synthetic, sumon2025cardiotabnet} and disease diagnosis~\cite{cassar2009chronic,ghaffar2023evaluation}, with an extensive impact~\cite{elstein2004origins}. 

At present, most existing studies primarily rely on single-modal medical data to perform disease diagnosis~\cite{chen2024ecg, hernandez2022synthetic, yao2024addressing, ansari2023deep}.
Although these single-modal approaches demonstrate certain effectiveness within their respective specific domains, their perception of physiology is often too limited to offer a holistic and comprehensive understanding of complex diseases. Taking diabetes as an example, its physiological manifestations are typically reflected across multiple modalities, such as altered heart rate variability in electrocardiograms (ECG), pulmonary complications observable in chest X-rays (CXR), and abnormal glucose or lipid levels revealed by laboratory blood tests (LAB)~\cite{lin2021deep}. Therefore, to address the risk of incomplete or inaccurate diagnoses resulting from reliance on a single modality, it is essential to integrate multimodal medical data for comprehensive and in-depth analysis of complex diseases~\cite{alcaraz2024cardiolab, steyaert2023multimodal}.

Consequently, there are a number of efforts that have emerged to explore leveraging multimodal medical data for disease diagnosis~\cite{kline2022multimodal,steyaert2023multimodal, venugopalan2021multimodal, abdelaziz2021alzheimer}. Nevertheless, these methods often make only simple and direct determinations about the presence or absence of a specific disease~\cite{gundapaneni2024deep, kumar2022deep}, but struggle with performing robust long-text diagnostic reasoning and generating interpretable clinical insights for multiple diseases, which severely hinders their practical application.

Recently, multimodal large language models (MLLMs)~\cite{zhang2023video, li2023llava, liu2023visual, liu2024mumu, tian2025audiox, wu2024next} have undergone rapid development and achieved impressive results in a variety of tasks, such as vision-language and audio-language tasks. They have demonstrated strong capabilities in integrating, generalizing, and reasoning across diverse data modalities, offering promising potential for generating interpretable disease diagnosis reports from medical data. Although several pioneering studies have made preliminary attempts to apply MLLMs in the medical field, such as for ECG analysis~\cite{zhao2024ecg, tian2024foundation} or medical image reporting~\cite{shentu2024cxr, liu2024bootstrapping, tanno2025collaboration} tasks, these works are still limited to single modalities (\textit{e.g.}, ECG, CXR) and remain at physiological-level understanding rather than disease-level reasoning. Therefore, an MLLM that can perceive and integrate heterogeneous multimodal medical data, thereby enabling interpretable multi-disease reasoning and diagnosis, remains a significant gap in current research.

Based on the observations above, we propose a novel MLLM framework, named \textbf{MedTVT-R1}, which leverages the complementarity and mutual corroboration of clinical multimodal medical data to enable reasoning and diagnosis of multiple complex diseases, with its advancements illustrated in Figure~\ref{fig:top_fig}.
To achieve this, we innovatively construct a well-curated instruction dataset, \textbf{MedTVT-QA}, which is the first attempt to simultaneously consider three heterogeneous modalities (\textit{i.e.}, ECG, CXR, and LAB), and provides corresponding question-answer (QA) pairs that not only cover physiological-level interpretations, but further explore disease-level diagnoses based on a Chain of Evidence (CoE) that fully leverages the complementarity and mutual corroboration among modalities, thereby establishing a solid foundation for MLLMs to progressively integrate multimodal medical data for physiological perception and multi-disease diagnosis.
Moreover, we introduce a modality perception layer (MPL) for MedTVT-R1, which can effectively capture the dependencies among different modalities and adaptively weight their contributions based on the relevance of each modality to specific diseases, thereby maximizing cross-modal interaction and information utilization.

To further unlock the potential of the constructed data, inspired by DeepSeek-R1~\cite{guo2025deepseek}, we also adopt Reinforcement Fine-Tuning (RFT) based on Group Relative Policy Optimization (GRPO) for post-training, with a dedicated Jaccard Reward function for multi-disease diagnostic scenarios, which substantially enhances the model’s reasoning capability.
Extensive experiments demonstrate the superiority of the proposed MedTVT-R1 in physiological-level understanding for each modality as well as in effectively leveraging multimodal features for disease-level diagnosis, which holds significant implications for applying MLLMs in clinical scenarios such as interpretable diagnostic report generation and complex comorbidity reasoning.
Our contributions are summarized as follows:
\begin{itemize}
    \item We introduce MedTVT-QA, the first medical instruction dataset that features heterogeneous modalities including ECG (Time Series), CXR (Visual Images), and LAB (Tabular Data), and offers QA pairs covering both physiological-level understanding and disease-level diagnosis with a Chain of Evidence, thus establishing a solid foundation for MLLMs to seamlessly integrate multimodal medical data for disease reasoning and diagnosis.
    \item We propose MedTVT-R1, a novel MLLM framework that fully leverages the complementarity and mutual corroboration among clinical multimodal data for interpretable diagnosis of complex comorbidities, with a modality perception layer that effectively captures inter-modal dependencies and adaptively weights the contribution of each modality.
    \item We employ a Reinforcement Fine-Tuning (RFT) strategy based on Group Relative Policy Optimization (GRPO) incorporating a dedicated Jaccard reward function to unlock data potential and enhance the model’s reasoning accuracy.
    \item Extensive experiments demonstrate that MedTVT-R1 achieves state-of-the-art performance in physiological representation understanding across various modalities and multimodal diagnosis and report generation for comorbidity.
\end{itemize}

\section{Related Work}
\textbf{MLLM for Medical Diagnosis.}
The application of Multimodal Large Language Models (MLLMs) in medical diagnosis has gained significant attention due to their ability to process and integrate diverse data modalities, such as text~\cite{li2025towards, lievin2024can, jin2024health, gallifant2025tripod, yuan2024continued}, images~\cite{irvin2019chexpert, lee2025cxr, lee2023llm, lu2024multimodal}, and tabular data~\cite{bisercic2023interpretable, huang2024critical}. Early works focused on single-modal approaches, such as text-based models for clinical note analysis~\cite{jin2024health, yuan2024continued, lievin2024can}, image-based models for radiology interpretation~\cite{lee2025cxr, irvin2019chexpert}, or ECG-based models for cardiac status analysis~\cite{zhao2024ecg, yu2023zero, lan2025gem, yang2025ecg}.  Although significant advancements have been made, existing research has yet to integrate \textbf{T}ime series data (e.g., Electrocardiograms), \textbf{V}isual data (e.g., chest X-rays), and \textbf{T}abular data (e.g., lab results) into a unified framework for comprehensive \textbf{Med}ical disease analysis and diagnosis. To bridge this gap, we introduce \textbf{MedTVT-R1}, a multimodal large language model designed to seamlessly combine CXR, ECG, and lab data through cross-modal interactions and contribution-aware operator, enabling accurate and interpretable disease diagnosis. 

\textbf{Reinforcement Learning with Verifiable Rewards.}
Group Relative Policy Optimization (GRPO)~\cite{guo2025deepseek}, unlike Proximal Policy Optimization (PPO)~\cite{yu2022surprising, schulman2017proximal} which estimates advantages through a reward model, approximates advantages by obtaining multiple samples from the LLM using the same prompt, with the advantage being the normalized reward for each response within its set of generated responses, achieving notable success in text-based tasks~\cite{shao2024deepseekmath, ramesh2024group, dao2025alphamaze} such as summarization and dialogue generation, as well as vision tasks~\cite{liu2025visual, tan2025reason} like image captioning.
Recently, GRPO has been applied to medical image analysis~\cite{lai2025med, pan2025medvlm}. However, it has not yet been utilized for multimodal tasks in the crucial area of multi-disease diagnosis, which requires the integration of text, images, time series, and tabular data. In this work, we are pioneering the application of GRPO with a newly designed reward function, the Jaccard Reward, to enhance the accuracy of multi-disease prediction. 

\section{Methodology}

\begin{figure}[h]
\centering
\vspace{-1.2em}
\includegraphics[width=0.96\linewidth]{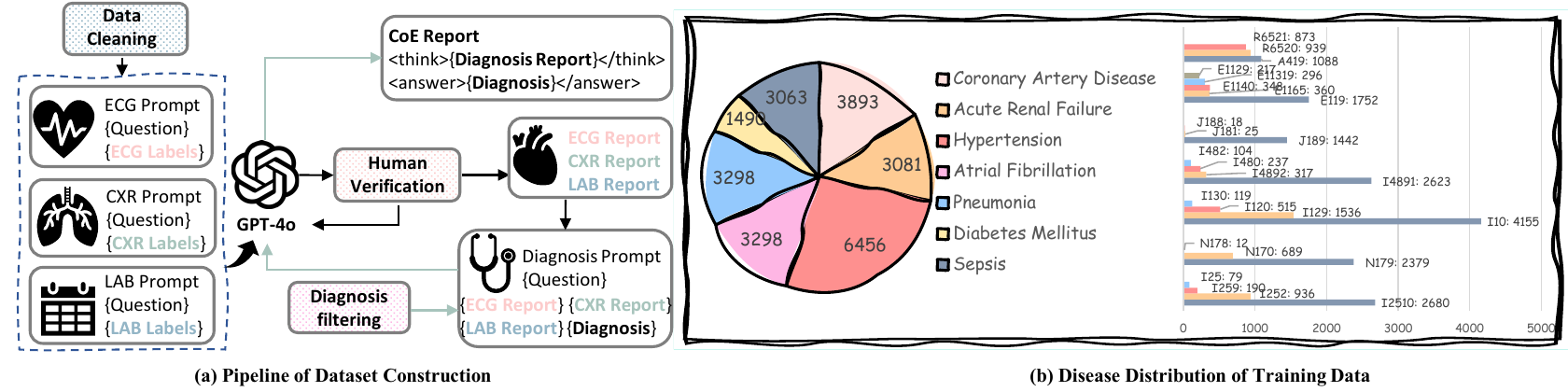}
\vspace{-0.6em}
  \caption{\small{MedTVT-QA dataset construction and disease distribution. (a) Pipeline of Dataset Construction: labels are refined to ensure consistency, prompts guide GPT-4o in generating verified physiological-level reports, which are combined with diagnostic labels to produce disease-level reports. Diagnostic labels are organized into seven primary categories with detailed subtypes. (b) Disease distribution of MedTVT-QA, with subtypes classified by ICD-10 codes. More details can be found in Appendix~\ref{sec:distrubution}.}}
\vspace{-1.2em}
\label{fig:data}
\end{figure} 

\subsection{MedTVT-QA}
\label{sec:MedTVT-QA}

To equip MLLMs with the ability to perform physiological understanding and disease diagnosis leveraging heterogeneous multimodal medical data, we collect raw medical data conforming to clinical temporal logic from the MIMIC-IV dataset, thanks to Symile~\cite{saporta2024contrasting}. Specifically, for each patient, we acquire electrocardiogram (ECG) readings and blood test results within the first 24 hours of hospital admission (from MIMIC-IV-ECG~\cite{gow2023mimic} and MIMIC-IV~\cite{johnson2020mimic}), as well as chest X-ray (CXR) images taken within 24 to 72 hours post-admission (from MIMIC-CXR-JPG~\cite{johnson2019mimic}), ultimately forming a total of 8,706 multimodal data combinations with consistent physiological temporal sequences, of which 8,331 for training and 375 for testing. Besides, the MIMIC-IV-ECG-EXT-ICD~\cite{MIMIC-IV-ECG-Ext-ICD} dataset provides emergency department and hospital discharge diagnoses, which are linked to the MIMIC-IV dataset. Based on these raw data, we progressively construct multimodal question-answer (QA) pairs from the perspectives of physiological-level representation analysis and disease-level diagnostic reasoning, with the entire process shown in Figure~\ref{fig:data} (a), which will be introduced in detail below.

\textbf{QA Pairs on Physiological-level Representation Analysis.}

To enable MLLMs to acquire a basic comprehension of the physiological meanings represented by each modality, \textit{i.e.}, ECG, CXR, and LAB, and to lay the groundwork for subsequent multi-disease reasoning and diagnosis based on multimodal data, we first construct QA pairs focused on physiological-level analysis for each modality. Specifically, taking the ECG modality as an example, we design a rigorous prompt with reference to physiological-level annotations, which consists of Role Setting, Task Description, Answer Guidance, and Answer Format to instruct GPT-4o~\cite{achiam2023gpt}, as outlined below:
\vspace{-0.4em}
\begin{tcolorbox}[colframe=gray, colback=gray!20, boxrule=0.3mm, top=0.7mm, bottom=0.7mm]
\tiny
\textbf{\textit{Role Setting}:} You are a renowned cardiologist with expertise in interpreting electrocardiograms (ECGs).  \\
\textbf{\textit{Task Description}:} The ECG analysis has yielded the following labels: \texttt{\{labels\}}. Based on these labels, please address the question: \texttt{\{question\}}.  \\
\textbf{\textit{Answer Guidance}:} Your response should incorporate all relevant labels, excluding any unrelated ones. Provide a synthesis of the labels, focusing on clinical significance. \\ 
\textbf{\textit{Answer Format}:}
Begin with a brief introduction to your analysis.
Provide detailed explanations for each specific ECG label.
Offer a concise summary.
\end{tcolorbox}
\vspace{-0.4em}
Here, \texttt{\{labels\}} refers to physiological-level annotations for the ECG data, such as Sinus Rhythm or Left Bundle Branch Block. \texttt{\{question\}} is randomly sampled from several carefully crafted question templates, such as `What information can be derived from the ECG?'. An example of a constructed QA pair that summarizes and analyzes the physiological representations encompassed in ECG is:
\vspace{-0.4em}
\begin{tcolorbox}[colframe=gray, colback=gray!20, boxrule=0.3mm, top=0.7mm, bottom=0.7mm]
\tiny
\textbf{\textit{Question}:} What information can be derived from the ECG <ecg>?  \\
\textbf{\textit{Answer}:} ECG Analysis: Upon evaluating the electrocardiogram (ECG), the following observations can be made: \\
1. Sinus Rhythm: The heart is under the control of the sinoatrial (SA) node, which is the natural pacemaker. This finding suggests that the electrical activity of the heart is originating from the SA node and progressing through the normal conduction pathway, leading to a regular rhythm. This is a foundational indicator of organized cardiac activity.\\
2. Left Bundle Branch Block (LBBB): There is evidence of delayed or disrupted conduction through the left bundle of the heart's electrical conduction system. This results in a characteristic pattern on the ECG, with a widened QRS complex (>120 ms) and specific morphologies in leads, such as a broad, notched appearance in V5-V6 and an absence of normal septal depolarization. LBBB may occur in conditions like ischemic heart disease, cardiomyopathy, or hypertension-induced structural changes, though it can also occasionally be found in asymptomatic individuals.\\
Summary: The heart demonstrates a sinus rhythm, indicating normal initiation of electrical activity. However, the presence of a left bundle branch block signifies an abnormality in electrical conduction, often associated with underlying cardiac pathology. As a whole, this ECG is classified as abnormal due to the conduction disturbance, warranting further investigation to assess structural or functional cardiac issues.
\end{tcolorbox}
\vspace{-0.4em}
Similar prompts are also applied to CXR and LAB data to organize the corresponding physiological-level QA pairs for each modality. It is worth noting that, for LAB data, we group 50 common laboratory indicators into seven categories according to their physiological significance to facilitate processing. The content generated above for physiological-level representation analysis undergoes manual review and revision by professionals to ensure its rationality and reliability. More details on physiological-level annotation and examples of QA pairs for each modality can be found in the Appendix~\ref{sec:prompts} to \ref{sec:QA_pairs}.

\textbf{QA Pairs on Disease-level Diagnostic Reasoning.}
Building upon the aforementioned completed physiological-level representation analysis for each modality, we further construct QA pairs that fully integrate information across modalities and conduct disease-level diagnostic reasoning, thereby enhancing the capability of MLLMs to handle complex multiple diseases. We focus on seven common and clinically significant diseases for which supporting evidence can be found in ECG, CXR, and LAB data, including \textit{Coronary Artery Disease, Acute Renal Failure, Hypertension, Atrial Fibrillation, Pneumonia, Diabetes Mellitus,} and \textit{Sepsis}, each of which contains several subtypes, with details in the Appendix~\ref{sec:distrubution}. The corresponding statistics are presented in Figure~\ref{fig:data} (b). We also employ a four-element prompt with reference to disease-level annotations to instruct GPT-4o, and compel its response to include a Chain of Evidence (CoE) to fully leverage the complementarity and mutual corroboration among modalities, thereby thoroughly extracting multimodal evidence for disease diagnosis, as follows:
\vspace{-0.4em}
\begin{tcolorbox}[colframe=gray, colback=gray!20, boxrule=0.3mm, top=0.7mm, bottom=0.7mm]
\tiny
\textbf{\textit{Role Setting}:} You are a renowned diagnostician with expertise in integrating ECG, CXR, and blood test results. \\
\textbf{\textit{Task Description}:} The following diagnostics have been provided: \\
$\bullet$ ECG Analysis: \texttt{\{ecg\_report\}} \\
$\bullet$ CXR Analysis: \texttt{\{cxr\_report\}} \\
$\bullet$ Blood Test Analysis: \texttt{\{blood\_test\_report\}} \\
$\bullet$ Diseases: \texttt{\{result\_diseases\}} \\
You need to pretend that the ECG, CXR, and blood test analyses are based on your interpretation of the raw data, and the final diagnosis is your synthesis of these three diagnostic methods, please address the question: \texttt{\{question\}} \\
\textbf{\textit{Answer Guidance}:} 
Please find definitive evidence from the ECG, CXR, and blood test results, leveraging the complementarity and mutual corroboration of these three modalities, to robustly prove the reasons why the patient has the diseases I provided. Your response must include every disease I provided, using the exact wording I provided, and you must not mention any diseases other than those I provided. Please make sure to provide evidence for these diagnoses! These are confirmed conditions.  \\
\textbf{\textit{Answer Format}:}
\texttt{<think>}\{Diagnostic evidence synthesized from the three modalities\}\texttt{</think>}\verb|\n| \texttt{<answer>}\{disease1; disease2; \dots\}\texttt{</answer>}
\end{tcolorbox}
\vspace{-0.4em}
Here, \texttt{\{ecg\_report\}}, \texttt{\{cxr\_report\}}, and \texttt{\{blood\_test\_report\}} respectively represent the physiological-level analyses of the three modalities. \texttt{\{result\_diseases\}} refers to the disease-level annotation of the sample. \texttt{\{question\}} is randomly sampled from several carefully crafted question templates, such as `Can you analyze my ECG, CXR and lab result to determine my probable conditions?'. The CoE is implemented by `Please find definitive evidence...'. The content obtained in this process is also reviewed by professionals to enhance its trustworthiness. An example of a QA pair that integrates multimodal information to mine evidence for multi-disease reasoning and diagnosis is shown in Figure~\ref{fig:framework}, and the complete version is provided in the Appendix~\ref{sec:performance}.

\subsection{MedTVT-R1}
\label{sec:framework}
\begin{figure}[t]
\centering
\includegraphics[width=1\linewidth]{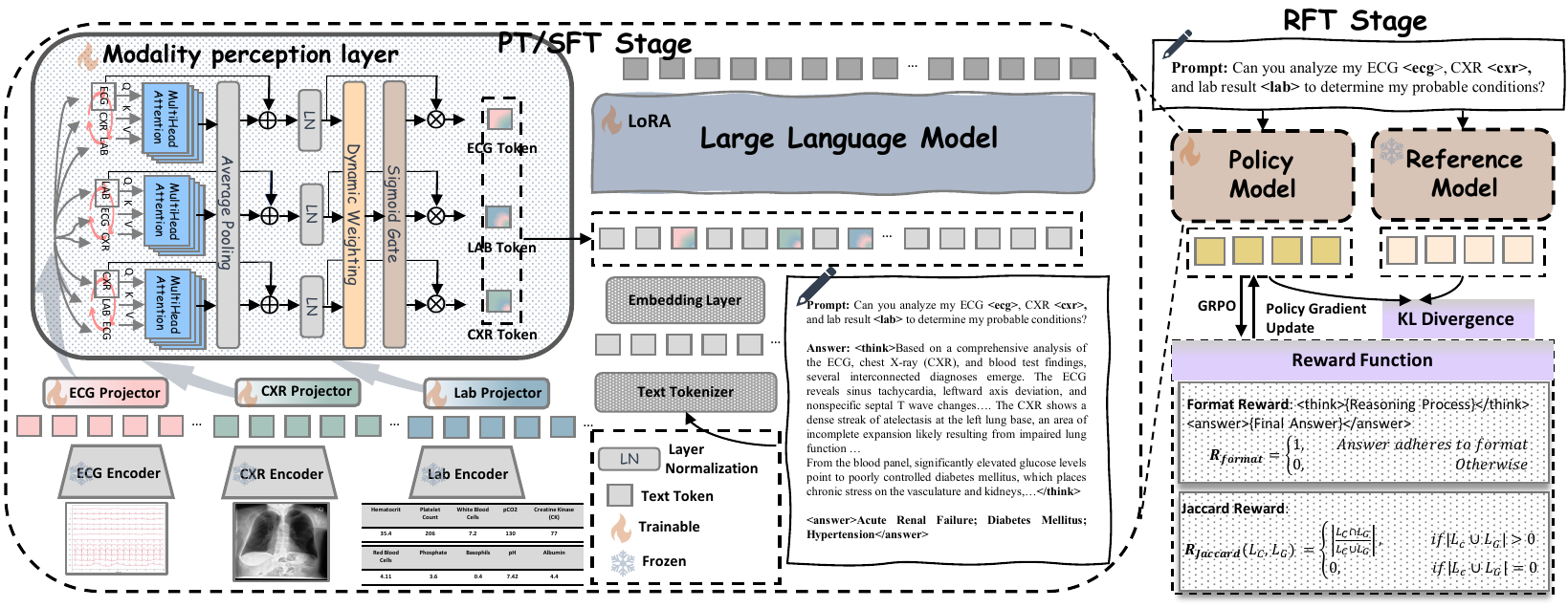}
\vspace{-1.9em}
  \caption{\small{Pipeline of MedTVT-R1. Pretraining processes ECG, CXR, and LAB data through encoders and projectors, combined with prompts, to train projectors and LLM’s LoRA for enhanced physiological understanding. The SFT stage adds a Modality Perception Layer for interaction and integration, refining disease analysis. The RFT stage applies GRPO, using the SFT-trained model for policy and inference, optimizing KL divergence and reward loss.  }}
\vspace{-1.5em}
\label{fig:framework}
\end{figure} 

Based on the meticulously constructed MedTVT-QA dataset described above, we propose MedTVT-R1, an MLLM framework capable of fully exploiting the complementarity and mutual corroboration of multimodal medical data for interpretable multi-disease reasoning and clinical diagnosis. In the following,  we will introduce the model architecture and training strategy of MedTVT-R1 in detail.

\subsubsection{Architecture}
The proposed MedTVT-R1 mainly consists of modality-specific encoders and projectors, a Modality Perception Layer (MPL), and a Large Language Model (LLM), with its overall architecture illustrated on the left side of Figure~\ref{fig:framework}. Given the raw data of ECG signals $\mathbf{X_E}\in\mathbb{R}^{N\times L}$, CXR images $\mathbf{X_C}\in\mathbb{R}^{C\times H \times W}$, and LAB tables $\mathbf{X_L}\in\mathbb{R}^{N'}$, they are first processed by their respective modality-specific encoders for feature extraction, and then the encoded features are fed into modality-specific projectors to a shared dimension $d$ for alignment and compatibility with the textual embedding space of the LLM, facilitating seamless integration between multimodal features and textual tokens; this process can be formulated as follows:
\begin{equation}
     \mathbf{Z_E} = g_\text{E}(f_\text{E}(\mathbf{X_E})) \in\mathbb{R}^{d}, \quad
     \mathbf{Z_C} = g_\text{C}(f_\text{C}(\mathbf{X_C})) \in\mathbb{R}^{d}, \quad
     \mathbf{Z_L} = g_\text{L}(f_\text{L}(\mathbf{X_L})) \in\mathbb{R}^{d},
\end{equation}
where $\mathbf{Z_{E/C/L}}$ denotes the projected multimodal features, and $f_\text{E/C/L}$ and $g_\text{E/C/L}$ represent the modality-specific encoders and projectors, respectively.

Subsequently, to enable efficient interaction and fusion among modalities, we introduce a Modality Perception Layer (MPL), which comprises a Cyclic Multi-Head Attention (CMHA) mechanism and a Contribution-Aware Operator (CAO). Specifically, the projected features $\mathbf{Z_E}$, $\mathbf{Z_C}$, and $\mathbf{Z_L}$ are first processed by the CMHA mechanism, in which each modality feature cyclically serves as the Query, Key, and Value to compute multi-head attention, enabling comprehensive capture of cross-modal dependencies and facilitating in-depth information exchange among ECG, CXR, and LAB features. After one round of cycling, the outputs are fused through average pooling, while a residual connection is employed to preserve modality-specific information. This process can be formulated as follows:
\begin{equation}
    \mathbf{F} = \text{AveragePooling}(\text{CMHA}(\mathbf{Z_E}, \mathbf{Z_C}, \mathbf{Z_L})), \quad \mathbf{M_{E/C/L}} = \mathbf{Z_{E/C/L}} + \mathbf{F},
\end{equation}
where $\mathbf{M_{E/C/L}}$ denotes the updated features of each modality, which encapsulate both modality-specific and modality-shared information. Recognizing that each modality contributes in varying degrees to the reasoning and diagnosis of various diseases, for example, ECG features are relatively more important for detecting Coronary Artery Disease, we design a Contribution-Aware Operator that adaptively assigns weights to the features of each modality based on the diagnostic context, which can be formulated as follows:
\begin{equation}
    \mathbf{T_E},\mathbf{T_C},\mathbf{T_L} = \sigma(h[\mathbf{M_E}:\mathbf{M_C}:\mathbf{M_L}])  \otimes (\mathbf{M_E},\mathbf{M_C},\mathbf{M_L}),
\end{equation}
where $[:]$ denotes the concatenation operation, $h$ is a learnable transformation matrix, $\sigma$ represents the Sigmoid activation, and $\otimes$ denotes element-wise multiplication. The final multimodal features $\mathbf{T_E}$, $\mathbf{T_C}$, and $\mathbf{T_L}$ are used to replace the placeholders \texttt{<ecg>}, \texttt{<cxr>}, and \texttt{<lab>} in the text tokens, which are obtained by processing the input prompt through the tokenizer and embedding layer. An example input prompt could be: `What illnesses might be indicated by the findings from my ECG \texttt{<ecg>}, CXR \texttt{<cxr>}, and blood work \texttt{<lab>}?'. The resulting sequence $\mathbf{T}_{\text{input}} = \{\mathbf{T}_{\text{Q}}, \mathbf{T}_{\text{E}}, \mathbf{T}_{\text{C}}, \mathbf{T}_{\text{L}}, \mathbf{T}_{\text{A}}\}$ is then fed into the LLM, where $\mathbf{T}_{\text{Q}}$ and $\mathbf{T}_{\text{A}}$ are derived from the QA pairs in the MedTVT-QA dataset.

\subsubsection{Training Strategy}
We employ a three-stage training strategy for MedTVT-R1, which includes Pre-training (PT), Supervised Fine-Tuning (SFT), and Reinforcement Fine-Tuning (RFT), to progressively enhance its ability to perceive the physiological representations of each modality and integrate multimodal information for interpretable multi-disease reasoning and diagnosis.

\textbf{Pre-training.}
With the aim of helping the model form an initial understanding and awareness of the physiological significance across all modalities, we first perform pre-training using physiological-level QA pairs from the MedTVT-QA dataset. During this stage, the projectors and the Low-Rank Adaptation (LoRA) modules embedded in the LLM are set as trainable, while the other components remain frozen. Notably, the MPL module is absent at this stage as no cross-modal interaction is involved. The optimization objective is to maximize the likelihood of generating the target response tokens, formalized as:
\begin{equation}
    \mathcal{L}_{\text{PT}} = -\mathbb{E}_{(\mathbf{T_Q}, \mathbf{T_{E/C/L}}, \mathbf{T_A}) \sim \mathcal{D}} 
\sum_{t=1}^{T} \log \pi_{\theta}(y_t \mid \mathbf{T_Q}, \mathbf{T_{E/C/L}}, y_{<t}),
\end{equation}
where $\pi_{\theta}(y_t \mid \cdot)$ denotes the conditional probability of generating the $t$-th token $y_t$, given the prompt, modality features, and the previously generated tokens $y_{<t}$.

\textbf{Supervised Fine-Tuning.}
With the pretrained model that already demonstrates a solid understanding of the physiological significance of each modality, we further conduct SFT based on disease-level QA pairs with CoE logic from the MedTVT-QA dataset to equip the model with the capability to synthesize multimodal representations and uncover the complementarity and mutual corroboration among modalities for multi-disease reasoning and diagnosis. During this stage, the MPL and the LoRA modules embedded in the LLM are set to be trainable while the other components remain frozen, and the optimization objective is similar to that of the pre-training stage, namely:
\begin{equation}
    \mathcal{L}_{\text{SFT}} = -\mathbb{E}_{(\mathbf{T_Q}, \mathbf{T_{E}}, \mathbf{T_{C}}, \mathbf{T_{L}}, \mathbf{T_A}) \sim \mathcal{D}} 
\sum_{t=1}^{T} \log \pi_{\theta}(y_t \mid \mathbf{T_Q}, \mathbf{T_{E}}, \mathbf{T_{C}}, \mathbf{T_{L}}, y_{<t}).
\end{equation}

\textbf{Reinforcement Fine-Tuning.}
To unlock the potential of the constructed dataset and boost the model’s reasoning performance, inspired by the advancements of DeepSeek-R1, we perform RFT using Group Relative Policy Optimization (GRPO) under the Reinforcement Learning with Verifiable Rewards (RLVR) framework. The training corpus and trainable components remain consistent with those in the SFT stage. The optimization objective can be formulated as:
\begin{equation}
    \max_{\pi_{\theta}} \mathbb{E}_{\mathbf{A} \sim \pi_{\theta}(\mathbf{Q})} \left[ R_{\text{RLVR}}(\mathbf{Q}, \mathbf{A}) \right]
    = \left[ R(\mathbf{Q}, \mathbf{A}) - \beta \text{KL}\left[\pi_{\theta}(\mathbf{A} \mid \mathbf{Q}) \, \| \, \pi_{\text{ref}}(\mathbf{A} \mid \mathbf{Q})\right] \right],
\end{equation}
where $\pi_{\theta}$ and $\pi_{\text{ref}}$ are the policy model and the reference model, respectively. $R$ is the verifiable reward function. $\text{KL}\left[\pi_{\theta}(\mathbf{A} \mid \mathbf{Q}) \, \| \, \pi_{\text{ref}}(\mathbf{A} \mid \mathbf{Q})\right]$ penalizes divergence from the reference policy $\pi_{\text{ref}}$, ensuring both correctness and alignment with prior knowledge. The hyperparameter $\beta$ controls the trade-off between reward maximization and policy regularization.

GRPO directly compares the relative quality of responses within a group without requiring an additional critic model. Specifically, given a question $\mathbf{Q}$, GRPO first generates $G$ candidate responses $\{o_1, o_2, \dots, o_G\}$ according to the current policy $\pi_{\theta_\text{old}}$, which are then assigned rewards $\{r_1, r_2, \dots, r_G\}$. The relative quality of these responses is calculated by normalizing the rewards using their mean and standard deviation. GRPO encourages the model to prioritize responses with higher relative rewards, fostering improved performance without requiring a separate critic.

The verifiable reward function $R$ consists of the Format Reward and the Jaccard Reward, \textit{i.e.}, $R = R_{\text{F}} + R_{\text{J}}$, ensuring both prediction accuracy and structural consistency. In line with DeepSeek-R1, the Format Reward $R_{\text{F}}$ is used to enforce the model's compliance with predefined formatting rules for the \texttt{<think>} and \texttt{<answer>} tags. The Jaccard Reward $R_J$ is a novel, meticulously designed reward function tailored for multi-disease diagnosis, which evaluates the alignment between the model’s predictions and the ground truth by leveraging the Jaccard similarity coefficient, thereby quantifying the overlap between the predicted and actual disease sets. Specifically, for each model completion and its corresponding ground truth, the disease sets within the \texttt{<answer>} tags are first extracted using regular expressions and denoted as $L_C = \{l_{c_1}, l_{c_2}, \dots, l_{c_m}\}$ and $L_G = \{l_{g_1}, l_{g_2}, \dots, l_{g_n}\}$, where $l_{c_i}$ and $l_{g_j}$ represent individual diseases in the predicted and ground truth sets, respectively. The Jaccard Reward $R_J$ is then computed as:
\begin{equation}
    R_{J}(L_C, L_G) = 
\begin{cases} 
\frac{|L_C \cap L_G|}{|L_C \cup L_G|}, & \text{if } |L_C \cup L_G| > 0, \\ 
0, & \text{if } |L_C \cup L_G| = 0.
\end{cases}
\end{equation}
When the union of the sets is not empty, the $R_{J}$ is determined by the ratio of the intersection size to the union size, thereby capturing the degree of overlap between the prediction and ground truth. If the union is empty, the $R_{J}$ is set to zero to ensure robustness against invalid or incomplete outputs. 
Therefore, the Jaccard reward encourages the model to generate outputs that are highly consistent with the ground truth labels, which effectively helps improve both the accuracy and reliability in multi-disease diagnosis scenarios.

\vspace{-0.5em}
\section{Experiments}

\begin{table*}[t!]
 \centering
 \resizebox{\textwidth}{!}{
\begin{tabular}{ll|cccc|cccc}
  \toprule
  \multirow{2}{*}{Method} & \multirow{2}{*}{LLM} & \multicolumn{4}{c|}{NLG} & \multicolumn{4}{c}{CE}\\
  \cmidrule(lr){3-6 } \cmidrule(lr){7-10} 
  ~ & ~  & BLEU & METEOR & ROUGE & BERT & PRECISION & RECALL & F1 SCORE & AUC   \\
  \midrule
  InternVL3-1B~\cite{zhu2025internvl3} & InternVL3-1B  & 0.0178 & 0.1884 &0.1265 & 0.8188 & 0.3333 & 0.1333 & 0.1904 &  0.5053 \\
  LLaVA-1.5-7B~\cite{liu2024improved} & Vicuna-7B & 0.0029 & 0.0809 &0.0681 & 0.7796  & 0.2495 & 0.1279 & 0.1691 & 0.5004\\
 LLaVA-One-Vision-7B~\cite{li2024llava} & Qwen2-7B & 0.0144 &  0.1618 & 0.1168 &0.8016 & 0.3120 &  0.1247&0.1782& 0.4975  \\
Qwen2.5-VL-3B-Instruct~\cite{bai2025qwen2} & Qwen2.5-3B-Instruct & 0.0218 & 0.2031 & 0.1331 & 0.8181 &  0.3493 & 0.1397 & 0.1995 & 0.5000 \\
Mini-InternVL-Chat-2B-V1-5~\cite{bai2025qwen2} & InternLM2-Chat-1.8B & 0.0092 & 0.1347 & 0.0959& 0.8008 &  0.2176 &  0.1343 & 0.1661 &  0.5015 \\
Molmo-7B-O-0924~\cite{deitke2024molmo} & OLMo-7B & 0.0155 & 0.1456 & 0.1070 & 0.8028 &  0.0295 & 0.0608 & 0.0398 & 0.5001 \\
Deepseek-VL-1.3B-Chat~\cite{lu2024deepseek} & Deepseek-1.3B-Chat & 0.0341 & 0.1756 & 0.1435 & 0.8128 &  0.2510 & 0.1278 & 0.1534 & 0.5021 \\
LLaVA-NeXT-8B~\cite{li2024llavanext-strong} & LLaMA3-8B & 0.0145 & 0.1532 & 0.1067 & 0.8145 &  0.2674 & 0.1294 & 0.1567 & 0.4987 \\
  \midrule
  {MedTVT-R1 w/o PT} & LLaMA3.2-1B & {0.1131 } & {0.3280} & { 0.2043} & {0.8599} & {0.4980} & {0.5208} & {0.4672} & {0.5851}\\
    {MedTVT-R1 w/o RFT} & LLaMA3.2-1B & {0.1325 } & {0.3499} & {0.2261} & {0.8660} & {0.5237} & {0.5783} & {0.4992} & {0.6242}\\
    \midrule
  \rowcolor{gray!30}
  {\bf MedTVT-R1} & LLaMA3.2-1B & {\bf 0.1353 } & {\bf 0.3536} & {\bf 0.2295} & {\bf 0.8652} & {\bf 0.5407} & {\bf 0.5908} & {\bf 0.5190} & {\bf 0.6554}\\
  
  \bottomrule
\end{tabular}}
\vspace{-0.3em}
 \caption{Comparison of MedTVT-R1 with various MLLMs and its variants on disease-level reasoning and diagnostic capabilities.  
 }
 \vspace{-1.6em}
 \label{tab:main_table}
\end{table*}

\subsection{Training Details and Metrics}
\textbf{Training Details.}
\label{sec:training_details}

We conduct all experiments on a server equipped with eight NVIDIA A800 80GB GPUs. For the LLM, we choose LLaMA 3.2-1B~\cite{grattafiori2024llama} and integrate the LoRA modules~\cite{hu2022lora} with a rank of 8 for fine-tuning. For the modality-specific encoders, we use the pre-trained weights from ECGFM-KED~\cite{tian2024foundation}, ViT-B/16~\cite{dosovitskiy2020image}, and Symile~\cite{saporta2024contrasting} for ECG, CXR, and LAB, respectively. All modality-specific projectors adopt the Dense block architecture from MuMu-LLaMA~\cite{liu2024mumu}, with the embedding dimension $d$ set to 2048. During training, the PT and SFT stages are each trained for 20 epochs, while the RFT stage is trained for 500 iterations using the open-source Trainer framework, with $G$ in GRPO set to 8.

\textbf{Metrics.} The effectiveness of multi-disease reasoning and diagnosis was evaluated from two perspectives. First, the descriptive accuracy of the generated diagnostic text was assessed using natural language generation (NLG) metrics, including BLEU, METEOR, ROUGE, and BERTScore. Second, the classification accuracy of multi-label disease categories in the responses was evaluated using clinical efficacy (CE) metrics, such as PRECISION, RECALL, F1 SCORE, and AUC.

\subsection{Quantitative Analysis}
\textbf{Disease-level Diagnostic Reasoning Results.} 
Since there is no multimodal large model capable of comprehensively analyzing ECG signals, medical images, and tabular data, we transform ECG signals into images and convert LAB tabular data into text, enabling feasible comparison with the existing leading MLLMs. Table~\ref{tab:main_table} presents the comparison results between our proposed model, MedTVT-R1, and eight state-of-the-art MLLMs, including InternVL3-1B~\cite{zhu2025internvl3}, LLaVA-1.5-7B~\cite{liu2024improved}, LLaVA-One-Vision-7B~\cite{li2024llava}, Qwen2.5-VL-3B-Instruct~\cite{bai2025qwen2}, Mini-InternVL-Chat-2B-V1-5~\cite{bai2025qwen2}, Molmo-7B-O-0924~\cite{deitke2024molmo}, Deepseek-VL-1.3B-Chat~\cite{lu2024deepseek}, and LLaVA-NeXT-8B~\cite{li2024llavanext-strong}. These MLLMs range in size from 1B to 8B and utilize various backbones, such as InternVL, Vicuna, OLMo, as well as Deepseek-VL, which incorporates reinforcement learning during training. All inference results were obtained using the open-source framework ModelScope SWIFT~\cite{zhao2025swift}.

The results demonstrate that MedTVT-R1 outperforms these leading models, excelling not only in natural language generation but also in clinical evaluation. This indicates the superior capability of MedTVT-R1 in both descriptive and diagnostic reasoning tasks in multi-disease scenarios. Furthermore, Table~\ref{tab:main_table} also presents ablation studies to investigate the impact of physiological-level pre-training and RFT-based post-training on the model's performance. The results show that removing either component leads to a noticeable decline in performance.
Specifically, the findings highlight two key insights: 
1) Physiological-level pre-training enables the model to acquire physiological knowledge across modalities in advance, which facilitates the differentiation and integration of multimodal physiological information during the subsequent SFT stage for disease-level diagnostic reasoning.
2) The RFT stage based on GRPO further unleashes the potential of the constructed data and enhances the model’s multi-disease diagnostic performance, enabling deeper and more effective cross-modal reasoning.

\begin{table*}[t!]
 \centering
 \resizebox{\textwidth}{!}{
\begin{tabular}{l|cccc|cccc|cccc}
  \toprule
  \multirow{2}{*}{Method}  & \multicolumn{4}{c|}{ECG-QA} & \multicolumn{4}{c|}{CXR-QA}  & \multicolumn{4}{c}{LAB-QA}\\
  \cmidrule(lr){2-5} \cmidrule(lr){6-9} \cmidrule(lr){10-13}
  ~  & BLEU & METEOR & ROUGE & BERT & BLEU & METEOR & ROUGE & BERT & BLEU & METEOR & ROUGE & BERT   \\
  \midrule
  InternVL3-1B~\cite{zhu2025internvl3}   & 0.0186 &  0.1795 & 0.1379 & 0.8282 & 0.0239 & 0.1827 &  0.1273 & 0.8309 & 0.0083 &  0.1234 &  0.0750 & 0.7750 \\
  LLaVA-1.5-7B~\cite{liu2024improved}  & 0.0055 & 0.1084 & 0.0866 & 0.8100  &  0.0034 & 0.0967 & 0.0812 & 0.8012 & 0.0170 & 0.1402 & 0.1133 & 0.7937\\
 LLaVA-One-Vision-7B~\cite{li2024llava}  & 0.0313 & 0.2263 &  0.1545 & 0.8322 & 0.0260 & 0.1877 & 0.1325 & 0.8214 & 0.0088 & 0.1362 & 0.0967 & 0.7883 \\
Qwen2.5-VL-3B-Instruct~\cite{bai2025qwen2}  &  0.0304 & 0.2483 & 0.1687 & 0.8418 &  0.0310 &  0.1798 &  0.1261 & 0.8230 & 0.0081 & 0.1129 & 0.0764 & 0.7832 \\
Mini-InternVL-Chat-2B-V1-5~\cite{bai2025qwen2}   & 0.0102 & 0.1336 & 0.0984 & 0.8112 & 0.0088 &0.1082 & 0.0825 & 0.8044 & 0.0085 &  0.1286 & 0.1118 & 0.7781 \\
  Molmo-7B-O-0924~\cite{deitke2024molmo}  & 0.0233 & 0.1949 & 0.1341 & 0.8305  &  0.0211 & 0.1813 & 0.1255 &0.8231 & 0.0091 & 0.1102 & 0.1120 & 0.7587\\
Deepseek-VL-1.3B-Chat~\cite{lu2024deepseek}  & 0.0240 & 0.1708 & 0.1162 & 0.8205 & 0.0298 & 0.1510 & 0.1312 & 0.8215 & 0.0118 & 0.0975 & 0.1184 & 0.7675 \\
LLaVA-NeXT-8B~\cite{li2024llavanext-strong}  & 0.0091 & 0.1412 & 0.1064 & 0.8009 &  0.0107 & 0.1305 & 0.1307 & 0.8199 & 0.0102 & 0.1057 & 0.1091 & 0.7623 \\
  \midrule
  \rowcolor{gray!30}
  {\bf MedTVT-R1} & {\bf 0.0831 } & {\bf 0.3044} & {\bf 0.2202} & {\bf  0.8650} & {\bf 0.0931} & {\bf 0.3073} & {\bf 0.2121} & {\bf 0.8673} & {\bf 0.1807} & {\bf0.3827} & {\bf 0.3081} & {\bf 0.8855}\\
  \bottomrule
\end{tabular}}
\vspace{-0.4em}
 \caption{Comparison of MedTVT-R1 with various MLLMs on physiological-level understanding and analysis capabilities. 
 }
 \vspace{-1.2em}
 \label{tab:singal-mode}
\end{table*}

\begin{figure}[t]
\centering
\includegraphics[width=1\linewidth]{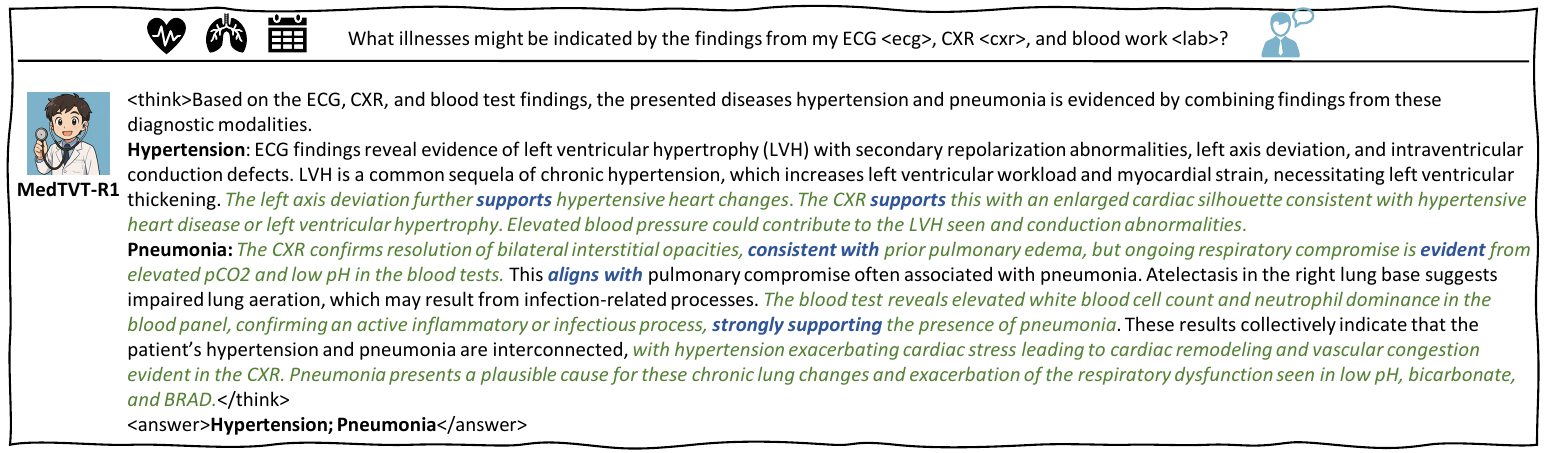}
\vspace{-1.5em}
  \caption{Inference example of MedTVT-R1: Blue highlights "evidence words," while green denotes critical information extracted from various modalities.}
\vspace{-1.8em}
\label{fig:QA}
\end{figure} 

\textbf{Physiological-level Understanding Results.} 
Similar to our analysis of Disease-level Diagnostic Reasoning, we evaluate MedTVT-R1 against eight MLLMs in terms of their single-modality understanding capabilities at the Physiological-level. The results in Table~\ref{tab:singal-mode} clearly demonstrate that our model outperforms all competitors. Notably, the Physiological-level representation analysis we designed is a challenging long-text generation task, requiring the generation of over 300 words per instance. Despite this complexity, MedTVT-R1 delivers outstanding results, further showcasing its advanced capabilities in handling lengthy and detailed outputs.
In a horizontal comparison, the understanding of LAB tabular data is superior to that of CXR images and ECG signals. This aligns with intuitive expectations, as LAB tabular data is inherently closer to a textual format, making it easier for large language models to process compared to the other two modalities.
These findings highlight the exceptional performance of MedTVT-R1 in long-text generation tasks and its ability to effectively comprehend and analyze physiological data across diverse modalities. Furthermore, the results emphasize the model’s robustness in both single-modality perception and multimodal reasoning, establishing MedTVT-R1 as a leading solution for complex medical data analysis.

\subsection{Qualitative Analysis}

MedTVT-R1 demonstrates a robust ability to integrate and analyze data from multiple modalities—CXR, ECG, and LAB tests—to arrive at comprehensive diagnoses. This integration allows for mutual corroboration among the modalities, enhancing diagnostic accuracy, as shown in Figure~\ref{fig:QA}. 
1) Multimodal Integration for Diagnosis: MedTVT-R1 effectively synthesizes information from CXR, ECG, and LAB data to diagnose conditions like hypertension and pneumonia. Each modality provides unique insights that collectively strengthen the diagnostic conclusion. For instance, ECG findings of left ventricular hypertrophy are corroborated by CXR observations of an enlarged cardiac silhouette, both indicative of hypertensive changes. Similarly, LAB results showing elevated white blood cell counts align with CXR evidence of interstitial opacities, supporting a pneumonia diagnosis. 
2) Evidence-Based Reasoning: The model frequently employs terms such as "support," "evident," and "aligns with," highlighting its capability to identify and utilize evidence from each modality to substantiate the final diagnosis. This approach demonstrates MedTVT-R1’s proficiency in extracting relevant features from each dataset, ensuring that the diagnostic reasoning is well-founded and comprehensive. For example, the alignment of elevated pCO2 and low pH with respiratory compromise underscores the model's ability to connect LAB findings with CXR results to confirm pneumonia.
Overall, MedTVT-R1's use of multimodal data not only enhances diagnostic precision but also ensures that each diagnosis is supported by a robust evidence base from all available modalities.
For more comparisons with the responses of other MLLMs, please refer to the appendix~\ref{sec:performance}.

\subsection{Ablation Study}

\begin{table}[t]
\tiny
\centering
\caption{The ablation of the  Cyclic Multi-Head Attention (CMHA) mechanism and Contribution-Aware Operator (CAO) in the  Modality Perception Layer (MPL), as well as the impact of modal missingness during the pre-training phase on the final results. Metrics include METEOR, ROUGE, RECALL, and F1 SCORE. Default settings are marked in \colorbox{baselinecolor}{gray}.}
\label{tab:combined}
\subfloat[
Ablation study of MPL.
\label{tab:abla1}
]{
\centering
\begin{minipage}{0.30\linewidth} 
\begin{center}
\begin{tabular}{x{10}x{10}|x{10}x{13}|x{10}x{13}}
\toprule[1pt]
\multicolumn{2}{c|}{\textbf{MPL}}  & \multicolumn{2}{c|}{\textbf{NLG}}  & \multicolumn{2}{c}{\textbf{CE}}\\
\cmidrule(lr){1-2} \cmidrule(lr){3-4} \cmidrule(lr){5-6}
\textbf{CMHA} & \textbf{CAO}   & \textbf{MET.} & \textbf{ROU.} & \textbf{REC. } & \textbf{F1} \\
\midrule
\xmark & \cmark  & 0.3455 & 0.2013 & 0.5733 & 0.4977  \\
\cmark & \xmark   & 0.3378 & 0.2145  & 0.5826 & 0.4867 \\
\rowcolor{gray!30}
\cmark & \cmark  & 0.3536  & 0.2295 & 0.5908 & 0.5190 \\
\bottomrule[1pt]
\end{tabular}
\end{center}
\end{minipage}
}
\hspace{8em}
\subfloat[
Ablation study of Modalities.
\label{tab:abla2}
]{
\centering
\begin{minipage}{0.36\linewidth} 
\begin{center}
\begin{tabular}{x{10}x{10}x{10}|x{10}x{13}|x{10}x{10}}
\toprule[1pt]
\multicolumn{3}{c|}{\textbf{Modality}}  & \multicolumn{2}{c|}{\textbf{NLG}}  & \multicolumn{2}{c}{\textbf{CE}}  \\
\cmidrule(lr){1-3} \cmidrule(lr){4-5} \cmidrule(lr){6-7}
\textbf{ECG} & \textbf{CXR} & \textbf{LAB}  & \textbf{MET.} & \textbf{ROU.} & \textbf{REC. } & \textbf{F1}  \\
\midrule
\xmark & \cmark & \cmark   &  0.3245 & 0.2058  &  0.5320 &  0.4739 \\
\cmark & \xmark & \cmark  &  0.3267  &  0.2019 &  0.5739 &  0.4869 \\
\cmark & \cmark & \xmark  &  0.3455 & 0.2218  &  0.5845 &  0.5023 \\
\rowcolor{gray!30}
\cmark & \cmark & \cmark  & 0.3536  & 0.2295 & 0.5908 & 0.5190\\
\bottomrule[1pt]
\end{tabular}
\end{center}
\end{minipage}
}
\vspace{-1em}
\end{table}

The ablation studies in Table~\ref{tab:combined} clearly demonstrate the effectiveness of both Cyclic Multi-Head Attention (CMHA) and Contribution-Aware Operator (CAO) in the MPL, as well as the necessity of complete modality integration during pre-training. The integration of CMHA and CAO significantly enhances the results, validating our initial intention to design these mechanisms to facilitate modality fusion and adjust the contribution levels of different modalities for various diseases. The presence of all three modalities (ECG, CXR, LAB) results in the best performance. Removing any single modality leads to reduced scores, with the absence of ECG showing the most significant drop in METEOR and ROUGE, possibly because many of the diseases are related to cardiac activity. This underscores the importance of multi-modal integration for optimal outcomes. 

\vspace{-0.5em}
\section{Conclusion}
\vspace{-0.5em}
\label{others}

In conclusion, the proposed MedTVT-R1 framework represents a significant advancement in the application of multimodal large language models (MLLMs) for medical diagnosis. By integrating the complementary strengths of ECG, CXR, and LAB data, MedTVT-R1 addresses the limitations of single-modal approaches and provides a more holistic understanding of complex diseases. The innovative MedTVT-QA dataset facilitates physiological perception and multi-disease diagnosis by leveraging a Chain of Evidence strategy. Additionally, the modality perception layer enhances cross-modal interactions, while Reinforcement Fine-Tuning with Group Relative Policy Optimization and the Jaccard Reward boosts precision and reliability in diagnosis capabilities. Extensive experiments validate MedTVT-R1's superior performance in both physiological-level understanding and disease-level diagnosis, highlighting its potential for practical clinical applications, such as interpretable diagnostic report generation and complex comorbidity reasoning. 


\bibliographystyle{abbrv}
\bibliography{reference.bib}

\begin{thebibliography}{10}

\bibitem{abdelaziz2021alzheimer}
M.~Abdelaziz, T.~Wang, and A.~Elazab.
\newblock Alzheimer’s disease diagnosis framework from incomplete multimodal data using convolutional neural networks.
\newblock {\em Journal of biomedical informatics}, 121:103863, 2021.

\bibitem{achiam2023gpt}
J.~Achiam, S.~Adler, S.~Agarwal, L.~Ahmad, I.~Akkaya, F.~L. Aleman, D.~Almeida, J.~Altenschmidt, S.~Altman, S.~Anadkat, et~al.
\newblock Gpt-4 technical report.
\newblock {\em arXiv preprint arXiv:2303.08774}, 2023.

\bibitem{alcaraz2024cardiolab}
J.~M.~L. Alcaraz and N.~Strodthoff.
\newblock Cardiolab: Laboratory values estimation and monitoring from electrocardiogram signals--a multimodal deep learning approach.
\newblock {\em arXiv preprint arXiv:2411.14886}, 2024.

\bibitem{ansari2023deep}
Y.~Ansari, O.~Mourad, K.~Qaraqe, and E.~Serpedin.
\newblock Deep learning for ecg arrhythmia detection and classification: an overview of progress for period 2017--2023.
\newblock {\em Frontiers in Physiology}, 14:1246746, 2023.

\bibitem{bai2025qwen2}
S.~Bai, K.~Chen, X.~Liu, J.~Wang, W.~Ge, S.~Song, K.~Dang, P.~Wang, S.~Wang, J.~Tang, et~al.
\newblock Qwen2. 5-vl technical report.
\newblock {\em arXiv preprint arXiv:2502.13923}, 2025.

\bibitem{bisercic2023interpretable}
A.~Bisercic, M.~Nikolic, M.~van~der Schaar, B.~Delibasic, P.~Lio, and A.~Petrovic.
\newblock Interpretable medical diagnostics with structured data extraction by large language models.
\newblock {\em arXiv preprint arXiv:2306.05052}, 2023.

\bibitem{ccalli2021deep}
E.~{\c{C}}all{\i}, E.~Sogancioglu, B.~van Ginneken, K.~G. van Leeuwen, and K.~Murphy.
\newblock Deep learning for chest x-ray analysis: A survey.
\newblock {\em Medical image analysis}, 72:102125, 2021.

\bibitem{cassar2009chronic}
A.~Cassar, D.~R. Holmes~Jr, C.~S. Rihal, and B.~J. Gersh.
\newblock Chronic coronary artery disease: diagnosis and management.
\newblock In {\em Mayo Clinic Proceedings}, volume~84, pages 1130--1146. Elsevier, 2009.

\bibitem{chen2024ecg}
X.~Chen, Y.~Niu, J.~Fan, L.~Lu, and H.~Fan.
\newblock Ecg-based intelligent model for coronary heart disease screening.
\newblock In {\em Proceedings of the 2024 11th International Conference on Biomedical and Bioinformatics Engineering}, pages 72--77, 2024.

\bibitem{dao2025alphamaze}
A.~Dao and D.~B. Vu.
\newblock Alphamaze: Enhancing large language models' spatial intelligence via grpo.
\newblock {\em arXiv preprint arXiv:2502.14669}, 2025.

\bibitem{deitke2024molmo}
M.~Deitke, C.~Clark, S.~Lee, R.~Tripathi, Y.~Yang, J.~S. Park, M.~Salehi, N.~Muennighoff, K.~Lo, L.~Soldaini, et~al.
\newblock Molmo and pixmo: Open weights and open data for state-of-the-art multimodal models.
\newblock {\em arXiv preprint arXiv:2409.17146}, 2024.

\bibitem{dosovitskiy2020image}
A.~Dosovitskiy, L.~Beyer, A.~Kolesnikov, D.~Weissenborn, X.~Zhai, T.~Unterthiner, M.~Dehghani, M.~Minderer, G.~Heigold, S.~Gelly, et~al.
\newblock An image is worth 16x16 words: Transformers for image recognition at scale.
\newblock {\em arXiv preprint arXiv:2010.11929}, 2020.

\bibitem{elstein2004origins}
A.~S. Elstein.
\newblock On the origins and development of evidence-based medicine and medical decision making.
\newblock {\em Inflammation research}, 53:S184--S189, 2004.

\bibitem{gallifant2025tripod}
J.~Gallifant, M.~Afshar, S.~Ameen, Y.~Aphinyanaphongs, S.~Chen, G.~Cacciamani, D.~Demner-Fushman, D.~Dligach, R.~Daneshjou, C.~Fernandes, et~al.
\newblock The tripod-llm reporting guideline for studies using large language models.
\newblock {\em Nature Medicine}, pages 1--10, 2025.

\bibitem{ghaffar2023evaluation}
N.~Ghaffar~Nia, E.~Kaplanoglu, and A.~Nasab.
\newblock Evaluation of artificial intelligence techniques in disease diagnosis and prediction.
\newblock {\em Discover Artificial Intelligence}, 3(1):5, 2023.

\bibitem{gow2023mimic}
B.~Gow, T.~Pollard, L.~A. Nathanson, A.~Johnson, B.~Moody, C.~Fernandes, N.~Greenbaum, J.~W. Waks, P.~Eslami, T.~Carbonati, et~al.
\newblock Mimic-iv-ecg: Diagnostic electrocardiogram matched subset.
\newblock {\em Type: dataset}, 6:13--14, 2023.

\bibitem{grattafiori2024llama}
A.~Grattafiori, A.~Dubey, A.~Jauhri, A.~Pandey, A.~Kadian, A.~Al-Dahle, A.~Letman, A.~Mathur, A.~Schelten, A.~Vaughan, et~al.
\newblock The llama 3 herd of models.
\newblock {\em arXiv preprint arXiv:2407.21783}, 2024.

\bibitem{gundapaneni2024deep}
S.~Gundapaneni, Z.~Zhi, and M.~Rodrigues.
\newblock Deep learning-based noninvasive screening of type 2 diabetes with chest x-ray images and electronic health records.
\newblock {\em arXiv preprint arXiv:2412.10955}, 2024.

\bibitem{guo2025deepseek}
D.~Guo, D.~Yang, H.~Zhang, J.~Song, R.~Zhang, R.~Xu, Q.~Zhu, S.~Ma, P.~Wang, X.~Bi, et~al.
\newblock Deepseek-r1: Incentivizing reasoning capability in llms via reinforcement learning.
\newblock {\em arXiv preprint arXiv:2501.12948}, 2025.

\bibitem{hernandez2022synthetic}
M.~Hernandez, G.~Epelde, A.~Alberdi, R.~Cilla, and D.~Rankin.
\newblock Synthetic data generation for tabular health records: A systematic review.
\newblock {\em Neurocomputing}, 493:28--45, 2022.

\bibitem{hu2022lora}
E.~J. Hu, Y.~Shen, P.~Wallis, Z.~Allen-Zhu, Y.~Li, S.~Wang, L.~Wang, W.~Chen, et~al.
\newblock Lora: Low-rank adaptation of large language models.
\newblock {\em ICLR}, 1(2):3, 2022.

\bibitem{huang2024critical}
J.~Huang, D.~M. Yang, R.~Rong, K.~Nezafati, C.~Treager, Z.~Chi, S.~Wang, X.~Cheng, Y.~Guo, L.~J. Klesse, et~al.
\newblock A critical assessment of using chatgpt for extracting structured data from clinical notes.
\newblock {\em npj Digital Medicine}, 7(1):106, 2024.

\bibitem{irvin2019chexpert}
J.~Irvin, P.~Rajpurkar, M.~Ko, Y.~Yu, S.~Ciurea-Ilcus, C.~Chute, H.~Marklund, B.~Haghgoo, R.~Ball, K.~Shpanskaya, et~al.
\newblock Chexpert: A large chest radiograph dataset with uncertainty labels and expert comparison.
\newblock In {\em Proceedings of the AAAI conference on artificial intelligence}, volume~33, pages 590--597, 2019.

\bibitem{jin2024health}
M.~Jin, Q.~Yu, C.~Zhang, D.~Shu, S.~Zhu, M.~Du, Y.~Zhang, and Y.~Meng.
\newblock Health-llm: Personalized retrieval-augmented disease prediction model.
\newblock {\em arXiv preprint arXiv:2402.00746}, 10, 2024.

\bibitem{johnson2020mimic}
A.~Johnson, L.~Bulgarelli, T.~Pollard, S.~Horng, L.~A. Celi, and R.~Mark.
\newblock Mimic-iv.
\newblock {\em PhysioNet. Available online at: https://physionet. org/content/mimiciv/1.0/(accessed August 23, 2021)}, pages 49--55, 2020.

\bibitem{johnson2019mimic}
A.~E. Johnson, T.~J. Pollard, N.~R. Greenbaum, M.~P. Lungren, C.-y. Deng, Y.~Peng, Z.~Lu, R.~G. Mark, S.~J. Berkowitz, and S.~Horng.
\newblock Mimic-cxr-jpg, a large publicly available database of labeled chest radiographs.
\newblock {\em arXiv preprint arXiv:1901.07042}, 2019.

\bibitem{kline2022multimodal}
A.~Kline, H.~Wang, Y.~Li, S.~Dennis, M.~Hutch, Z.~Xu, F.~Wang, F.~Cheng, and Y.~Luo.
\newblock Multimodal machine learning in precision health: A scoping review.
\newblock {\em npj Digital Medicine}, 5(1):171, 2022.

\bibitem{kumar2022deep}
A.~Kumar.
\newblock Deep learning for multi-modal medical imaging fusion: Enhancing diagnostic accuracy in complex disease detection.
\newblock {\em Int J Eng Technol Res Manag}, 6(11):183, 2022.

\bibitem{lai2025med}
Y.~Lai, J.~Zhong, M.~Li, S.~Zhao, and X.~Yang.
\newblock Med-r1: Reinforcement learning for generalizable medical reasoning in vision-language models.
\newblock {\em arXiv preprint arXiv:2503.13939}, 2025.

\bibitem{lan2025gem}
X.~Lan, F.~Wu, K.~He, Q.~Zhao, S.~Hong, and M.~Feng.
\newblock Gem: Empowering mllm for grounded ecg understanding with time series and images.
\newblock {\em arXiv preprint arXiv:2503.06073}, 2025.

\bibitem{lee2023llm}
S.~Lee, W.~J. Kim, J.~Chang, and J.~C. Ye.
\newblock Llm-cxr: instruction-finetuned llm for cxr image understanding and generation.
\newblock {\em arXiv preprint arXiv:2305.11490}, 2023.

\bibitem{lee2025cxr}
S.~Lee, J.~Youn, H.~Kim, M.~Kim, and S.~H. Yoon.
\newblock Cxr-llava: a multimodal large language model for interpreting chest x-ray images.
\newblock {\em European Radiology}, pages 1--13, 2025.

\bibitem{li2024llavanext-strong}
B.~Li, K.~Zhang, H.~Zhang, D.~Guo, R.~Zhang, F.~Li, Y.~Zhang, Z.~Liu, and C.~Li.
\newblock Llava-next: Stronger llms supercharge multimodal capabilities in the wild, May 2024.

\bibitem{li2024llava}
B.~Li, Y.~Zhang, D.~Guo, R.~Zhang, F.~Li, H.~Zhang, K.~Zhang, P.~Zhang, Y.~Li, Z.~Liu, et~al.
\newblock Llava-onevision: Easy visual task transfer.
\newblock {\em arXiv preprint arXiv:2408.03326}, 2024.

\bibitem{li2023llava}
C.~Li, C.~Wong, S.~Zhang, N.~Usuyama, H.~Liu, J.~Yang, T.~Naumann, H.~Poon, and J.~Gao.
\newblock Llava-med: Training a large language-and-vision assistant for biomedicine in one day.
\newblock {\em Advances in Neural Information Processing Systems}, 36:28541--28564, 2023.

\bibitem{li2025towards}
C.-Y. Li, K.-J. Chang, C.-F. Yang, H.-Y. Wu, W.~Chen, H.~Bansal, L.~Chen, Y.-P. Yang, Y.-C. Chen, S.-P. Chen, et~al.
\newblock Towards a holistic framework for multimodal llm in 3d brain ct radiology report generation.
\newblock {\em Nature Communications}, 16(1):2258, 2025.

\bibitem{lievin2024can}
V.~Li{\'e}vin, C.~E. Hother, A.~G. Motzfeldt, and O.~Winther.
\newblock Can large language models reason about medical questions?
\newblock {\em Patterns}, 5(3), 2024.

\bibitem{lin2021deep}
C.-S. Lin, Y.-T. Lee, W.-H. Fang, Y.-S. Lou, F.-C. Kuo, C.-C. Lee, and C.~Lin.
\newblock Deep learning algorithm for management of diabetes mellitus via electrocardiogram-based glycated hemoglobin (ecg-hba1c): a retrospective cohort study.
\newblock {\em Journal of Personalized Medicine}, 11(8):725, 2021.

\bibitem{liu2024bootstrapping}
C.~Liu, Y.~Tian, W.~Chen, Y.~Song, and Y.~Zhang.
\newblock Bootstrapping large language models for radiology report generation.
\newblock In {\em Proceedings of the AAAI Conference on Artificial Intelligence}, volume~38, pages 18635--18643, 2024.

\bibitem{liu2024improved}
H.~Liu, C.~Li, Y.~Li, and Y.~J. Lee.
\newblock Improved baselines with visual instruction tuning.
\newblock In {\em Proceedings of the IEEE/CVF Conference on Computer Vision and Pattern Recognition}, pages 26296--26306, 2024.

\bibitem{liu2023visual}
H.~Liu, C.~Li, Q.~Wu, and Y.~J. Lee.
\newblock Visual instruction tuning.
\newblock {\em Advances in neural information processing systems}, 36:34892--34916, 2023.

\bibitem{liu2024mumu}
S.~Liu, A.~S. Hussain, Q.~Wu, C.~Sun, and Y.~Shan.
\newblock Mumu-llama: Multi-modal music understanding and generation via large language models.
\newblock {\em arXiv preprint arXiv:2412.06660}, 2024.

\bibitem{liu2021deep}
X.~Liu, H.~Wang, Z.~Li, and L.~Qin.
\newblock Deep learning in ecg diagnosis: A review.
\newblock {\em Knowledge-Based Systems}, 227:107187, 2021.

\bibitem{liu2025visual}
Z.~Liu, Z.~Sun, Y.~Zang, X.~Dong, Y.~Cao, H.~Duan, D.~Lin, and J.~Wang.
\newblock Visual-rft: Visual reinforcement fine-tuning.
\newblock {\em arXiv preprint arXiv:2503.01785}, 2025.

\bibitem{lu2024deepseek}
H.~Lu, W.~Liu, B.~Zhang, B.~Wang, K.~Dong, B.~Liu, J.~Sun, T.~Ren, Z.~Li, H.~Yang, et~al.
\newblock Deepseek-vl: towards real-world vision-language understanding.
\newblock {\em arXiv preprint arXiv:2403.05525}, 2024.

\bibitem{lu2024multimodal}
M.~Y. Lu, B.~Chen, D.~F. Williamson, R.~J. Chen, M.~Zhao, A.~K. Chow, K.~Ikemura, A.~Kim, D.~Pouli, A.~Patel, et~al.
\newblock A multimodal generative ai copilot for human pathology.
\newblock {\em Nature}, 634(8033):466--473, 2024.

\bibitem{pan2025medvlm}
J.~Pan, C.~Liu, J.~Wu, F.~Liu, J.~Zhu, H.~B. Li, C.~Chen, C.~Ouyang, and D.~Rueckert.
\newblock Medvlm-r1: Incentivizing medical reasoning capability of vision-language models (vlms) via reinforcement learning.
\newblock {\em arXiv preprint arXiv:2502.19634}, 2025.

\bibitem{ramesh2024group}
S.~S. Ramesh, Y.~Hu, I.~Chaimalas, V.~Mehta, P.~G. Sessa, H.~Bou~Ammar, and I.~Bogunovic.
\newblock Group robust preference optimization in reward-free rlhf.
\newblock {\em Advances in Neural Information Processing Systems}, 37:37100--37137, 2024.

\bibitem{saporta2024contrasting}
A.~Saporta, A.~M. Puli, M.~Goldstein, and R.~Ranganath.
\newblock Contrasting with symile: Simple model-agnostic representation learning for unlimited modalities.
\newblock {\em Advances in Neural Information Processing Systems}, 37:56919--56957, 2024.

\bibitem{schulman2017proximal}
J.~Schulman, F.~Wolski, P.~Dhariwal, A.~Radford, and O.~Klimov.
\newblock Proximal policy optimization algorithms.
\newblock {\em arXiv preprint arXiv:1707.06347}, 2017.

\bibitem{shao2024deepseekmath}
Z.~Shao, P.~Wang, Q.~Zhu, R.~Xu, J.~Song, X.~Bi, H.~Zhang, M.~Zhang, Y.~Li, Y.~Wu, et~al.
\newblock Deepseekmath: Pushing the limits of mathematical reasoning in open language models.
\newblock {\em arXiv preprint arXiv:2402.03300}, 2024.

\bibitem{shentu2024cxr}
J.~Shentu and N.~Al~Moubayed.
\newblock Cxr-irgen: an integrated vision and language model for the generation of clinically accurate chest x-ray image-report pairs.
\newblock In {\em Proceedings of the IEEE/CVF Winter Conference on Applications of Computer Vision}, pages 5212--5221, 2024.

\bibitem{steyaert2023multimodal}
S.~Steyaert, M.~Pizurica, D.~Nagaraj, P.~Khandelwal, T.~Hernandez-Boussard, A.~J. Gentles, and O.~Gevaert.
\newblock Multimodal data fusion for cancer biomarker discovery with deep learning.
\newblock {\em Nature machine intelligence}, 5(4):351--362, 2023.

\bibitem{MIMIC-IV-ECG-Ext-ICD}
N.~Strodthoff, J.~M. Lopez~Alcaraz, and W.~Haverkamp.
\newblock Mimic-iv-ecg-ext-icd: Diagnostic labels for mimic-iv-ecg (version 1.0.1).
\newblock {\em PhysioNet}, 2024.

\bibitem{sumon2025cardiotabnet}
M.~S.~I. Sumon, M.~S.~B. Islam, M.~S. Rahman, M.~S.~A. Hossain, A.~Khandakar, A.~Hasan, M.~Murugappan, and M.~E. Chowdhury.
\newblock Cardiotabnet: A novel hybrid transformer model for heart disease prediction using tabular medical data.
\newblock {\em arXiv preprint arXiv:2503.17664}, 2025.

\bibitem{tan2025reason}
H.~Tan, Y.~Ji, X.~Hao, M.~Lin, P.~Wang, Z.~Wang, and S.~Zhang.
\newblock Reason-rft: Reinforcement fine-tuning for visual reasoning.
\newblock {\em arXiv preprint arXiv:2503.20752}, 2025.

\bibitem{tanno2025collaboration}
R.~Tanno, D.~G. Barrett, A.~Sellergren, S.~Ghaisas, S.~Dathathri, A.~See, J.~Welbl, C.~Lau, T.~Tu, S.~Azizi, et~al.
\newblock Collaboration between clinicians and vision--language models in radiology report generation.
\newblock {\em Nature Medicine}, 31(2):599--608, 2025.

\bibitem{tian2024foundation}
Y.~Tian, Z.~Li, Y.~Jin, M.~Wang, X.~Wei, L.~Zhao, Y.~Liu, J.~Liu, and C.~Liu.
\newblock Foundation model of ecg diagnosis: Diagnostics and explanations of any form and rhythm on ecg.
\newblock {\em Cell Reports Medicine}, 5(12), 2024.

\bibitem{tian2025audiox}
Z.~Tian, Y.~Jin, Z.~Liu, R.~Yuan, X.~Tan, Q.~Chen, W.~Xue, and Y.~Guo.
\newblock Audiox: Diffusion transformer for anything-to-audio generation.
\newblock {\em arXiv preprint arXiv:2503.10522}, 2025.

\bibitem{venugopalan2021multimodal}
J.~Venugopalan, L.~Tong, H.~R. Hassanzadeh, and M.~D. Wang.
\newblock Multimodal deep learning models for early detection of alzheimer’s disease stage.
\newblock {\em Scientific reports}, 11(1):3254, 2021.

\bibitem{wu2024next}
S.~Wu, H.~Fei, L.~Qu, W.~Ji, and T.-S. Chua.
\newblock Next-gpt: Any-to-any multimodal llm.
\newblock In {\em Forty-first International Conference on Machine Learning}, 2024.

\bibitem{yang2025ecg}
K.~Yang, M.~Hong, J.~Zhang, Y.~Luo, S.~Zhao, O.~Zhang, X.~Yu, J.~Zhou, L.~Yang, P.~Zhang, et~al.
\newblock Ecg-lm: Understanding electrocardiogram with a large language model.
\newblock {\em Health Data Science}, 5:0221, 2025.

\bibitem{yao2024addressing}
W.~Yao, C.~Liu, K.~Yin, W.~Cheung, and J.~Qin.
\newblock Addressing asynchronicity in clinical multimodal fusion via individualized chest x-ray generation.
\newblock {\em Advances in Neural Information Processing Systems}, 37:29001--29028, 2024.

\bibitem{yu2022surprising}
C.~Yu, A.~Velu, E.~Vinitsky, J.~Gao, Y.~Wang, A.~Bayen, and Y.~Wu.
\newblock The surprising effectiveness of ppo in cooperative multi-agent games.
\newblock {\em Advances in neural information processing systems}, 35:24611--24624, 2022.

\bibitem{yu2023zero}
H.~Yu, P.~Guo, and A.~Sano.
\newblock Zero-shot ecg diagnosis with large language models and retrieval-augmented generation.
\newblock In {\em Machine learning for health (ML4H)}, pages 650--663. PMLR, 2023.

\bibitem{yuan2024continued}
D.~Yuan, E.~Rastogi, G.~Naik, S.~P. Rajagopal, S.~Goyal, F.~Zhao, B.~Chintagunta, and J.~Ward.
\newblock A continued pretrained llm approach for automatic medical note generation.
\newblock {\em arXiv preprint arXiv:2403.09057}, 2024.

\bibitem{zhang2023video}
H.~Zhang, X.~Li, and L.~Bing.
\newblock Video-llama: An instruction-tuned audio-visual language model for video understanding.
\newblock {\em arXiv preprint arXiv:2306.02858}, 2023.

\bibitem{zhao2025swift}
Y.~Zhao, J.~Huang, J.~Hu, X.~Wang, Y.~Mao, D.~Zhang, Z.~Jiang, Z.~Wu, B.~Ai, A.~Wang, et~al.
\newblock Swift: a scalable lightweight infrastructure for fine-tuning.
\newblock In {\em Proceedings of the AAAI Conference on Artificial Intelligence}, volume~39, pages 29733--29735, 2025.

\bibitem{zhao2024ecg}
Y.~Zhao, T.~Zhang, X.~Wang, P.~Han, T.~Chen, L.~Huang, Y.~Jin, and J.~Kang.
\newblock Ecg-chat: A large ecg-language model for cardiac disease diagnosis.
\newblock {\em arXiv preprint arXiv:2408.08849}, 2024.

\bibitem{zhu2025internvl3}
J.~Zhu, W.~Wang, Z.~Chen, Z.~Liu, S.~Ye, L.~Gu, Y.~Duan, H.~Tian, W.~Su, J.~Shao, et~al.
\newblock Internvl3: Exploring advanced training and test-time recipes for open-source multimodal models.
\newblock {\em arXiv preprint arXiv:2504.10479}, 2025.

\end{thebibliography}


\newpage
\appendix

\section{Details About the Prompts of MedTVT-QA's Construction}
\label{sec:prompts}
This section presents the detailed prompts used in constructing the MedTVT-QA dataset. 

\begin{tcolorbox}[colframe=gray, colback=gray!20, boxrule=0.5mm, title={ECG-QA Prompt}]
\scriptsize
\textbf{\textit{Role Setting}:} You are a renowned cardiologist with expertise in interpreting electrocardiograms (ECGs).  \\
\textbf{\textit{Task Description}:} The ECG analysis has yielded the following labels: \texttt{\{labels\}}. Based on these labels, please address the question: \texttt{\{question\}}.  \\
\textbf{\textit{Answer Guidance}:} Your response should incorporate all relevant labels, excluding any unrelated ones. Provide a synthesis of the labels, focusing on clinical significance. \\ 
\textbf{\textit{Answer Format}:}
Begin with a brief introduction to your analysis.
Provide detailed explanations for each specific ECG label.
Offer a concise summary.
\end{tcolorbox}

\begin{tcolorbox}[colframe=gray, colback=gray!20, boxrule=0.5mm, title={CXR-QA Prompt}]
\scriptsize
\textbf{\textit{Role Setting}:} You are a radiology expert with expertise in interpreting chest X-ray image.\\
\textbf{\textit{Task Description}:} The chest X-ray report is given \texttt{\{report\}}
Base on the given chest X-ray report, answer the question \texttt{\{question\}} \\
\textbf{\textit{Answer Guidance}:} Describe the overall condition of the lungs, heart, and chest cavity in the image.
Identify and explain any abnormal findings such as shadows, opacities, effusions, or masses.
Provide possible diagnoses.\\ 
\end{tcolorbox}

\begin{tcolorbox}[colframe=gray, colback=gray!20, boxrule=0.5mm, title={LAB-QA Prompt}]
\scriptsize
\textbf{\textit{Role Setting}:} Please analyze this set of blood test data as a medical professional.  \\
\textbf{\textit{Task Description}:} This is the question: \texttt{\{question\}}
The following are the lab data:
"Hematocrit": \texttt{\{data[0]\}}; "Platelet Count": \texttt{\{data[1]\}}; "Creatinine": \texttt{\{data[2]\}}; "Potassium": \texttt{\{data[3]\}}; "Hemoglobin": \texttt{\{data[4]\}}; "White Blood Cells": \texttt{\{data[5]\}}; "MCHC": \texttt{\{data[6]\}}; "Red Blood Cells": \texttt{\{data[7]\}}; "MCV": \texttt{\{data[8]\}}; "MCH": \texttt{\{data[9]\}}; "RDW": \texttt{\{data[10]\}}; "Urea Nitrogen": \texttt{\{data[11]\}}; "Sodium": \texttt{\{data[12]\}}; "Chloride": \texttt{\{data[13]\}}; "Bicarbonate": \texttt{\{data[14]\}}; "Anion Gap": \texttt{\{data[15]\}}; "Glucose": \texttt{\{data[16]\}}; "Magnesium": \texttt{\{data[17]\}}; "Calcium, Total": \texttt{\{data[18]\}}; "Phosphate": \texttt{\{data[19]\}}; "INR(PT)": \texttt{\{data[20]\}}; "PT": \texttt{\{data[21]\}}; "PTT": \texttt{\{data[22]\}}; "Basophils": \texttt{\{data[23]\}}; "Neutrophils": \texttt{\{data[24]\}}; "Monocytes": \texttt{\{data[25]\}}; "Eosinophils": \texttt{\{data[26]\}}; "Lymphocytes": \texttt{\{data[27]\}}; "RDW-SD": \texttt{\{data[28]\}}; "H": \texttt{\{data[29]\}}; "L": \texttt{\{data[30]\}}; "I": \texttt{\{data[31]\}}; "Alanine Aminotransferase (ALT)": \texttt{\{data[32]\}}; "Asparate Aminotransferase (AST)": \texttt{\{data[33]\}}; "Lactate": \texttt{\{data[34]\}}; "Alkaline Phosphatase": \texttt{\{data[35]\}}; "Bilirubin, Total": \texttt{\{data[36]\}}; "pH": \texttt{\{data[37]\}}; "Albumin": \texttt{\{data[38]\}}; "Base Excess": \texttt{\{data[39]\}}; "pO2": \texttt{\{data[40]\}}
"Calculated Total CO2": \texttt{\{data[41]\}}; "pCO2": \texttt{\{data[42]\}}; "Absolute Neutrophil Count": \texttt{\{data[43]\}}; "Absolute Eosinophil Count": \texttt{\{data[44]\}}; "Absolute Monocyte Count": \texttt{\{data[45]\}}; "Absolute Basophil Count": \texttt{\{data[46]\}}; "Absolute Lymphocyte Count": \texttt{\{data[47]\}}; "Creatine Kinase (CK)": \texttt{\{data[48]\}}
"Immature Granulocytes": \texttt{\{data[49]\}}\\
\textbf{\textit{Answer Guidance}:} These data comprise 50 different indicators, categorized into seven main classes: routine blood indicators, electrolyte and metabolic indicators, renal function indicators, liver function indicators, acid-base balance and gas exchange, coagulation function indicators, and other indicators. \\ 
\textbf{\textit{Answer Format}:}
Begin with a brief introduction to your analysis.\\
\textbf{routine blood indicators}: explanation\\
\textbf{electrolyte and metabolic indicators}: explanation\\
\textbf{renal function indicators}: explanation\\
\textbf{liver function indicators}: explanation\\
\textbf{acid-base balance and gas exchange}: explanation\\
\textbf{coagulation function indicators}: explanation\\
\textbf{other indicators}: explanation\\
Finally, offer a concise summary.\\
\end{tcolorbox}

\begin{tcolorbox}[colframe=gray, colback=gray!20, boxrule=0.5mm, title={Disease-QA Prompt}]
\scriptsize
\textbf{\textit{Role Setting}:} You are a renowned diagnostician with expertise in integrating ECG, CXR, and blood test results. \\
\textbf{\textit{Task Description}:} The following diagnostics have been provided: \\
$\bullet$ ECG Analysis: \texttt{\{ecg\_report\}} \\
$\bullet$ CXR Analysis: \texttt{\{cxr\_report\}} \\
$\bullet$ Blood Test Analysis: \texttt{\{blood\_test\_report\}} \\
$\bullet$ Diseases: \texttt{\{result\_diseases\}} \\
You need to pretend that the ECG, CXR, and blood test analyses are based on your interpretation of the raw data, and the final diagnosis is your synthesis of these three diagnostic methods, please address the question: \texttt{\{question\}} \\
\textbf{\textit{Answer Guidance}:} 
Please find definitive evidence from the ECG, CXR, and blood test results, leveraging the complementarity and mutual corroboration of these three modalities, to robustly prove the reasons why the patient has the diseases I provided. Your response must include every disease I provided, using
the exact wording I provided, and you must not mention any diseases other than those I provided. Please make sure to provide evidence for these diagnoses!
These are confirmed conditions. \\
\textbf{\textit{Answer Format}:}
\texttt{<think>}\{Diagnostic evidence synthesized from the three modalities\}\texttt{</think>}\verb|\n| \texttt{<answer>}\{disease1; disease2; \dots\}\texttt{</answer>}
\end{tcolorbox}

\section{Label Distribution of MedTVT-QA}
\label{sec:distrubution}
When constructing the physiology-level ECG-QA dataset, we filtered out invalid ECG labels to ensure that the final labels align with morphology descriptions at the physiological level. Additionally, we conducted a detailed statistical analysis of the labels in the ECG-QA training data. As shown in Table~\ref{tab:ecg_label}, it presents ECG labels with occurrences greater than 100 along with their respective counts.

\begingroup
\small
\begin{longtable}{|l|r|}
\caption{ECG Labels and Counts (>100) in ECG-QA.} \\
\hline
\label{tab:ecg_label}
\textbf{Label} & \textbf{Count} \\
\hline
\endfirsthead
\hline
\textbf{Label} & \textbf{Count} \\
\hline
\endhead
\hline
\endfoot
\hline
\endlastfoot
sinus rhythm with 1st degree a-v block & 140 \\
sinus rhythm & 4033 \\
atrial fibrillation & 761 \\
sinus tachycardia & 1565 \\
consider acute st elevation mi & 161 \\
atrial fibrillation with rapid ventricular response & 224 \\
age not entered, assumed to be 50 years old for purpose of ecg interpretation & 328 \\
sinus bradycardia & 402 \\
sinus rhythm with pac(s) & 132 \\
sinus rhythm with borderline 1st degree a-v block & 121 \\
pacemaker rhythm - no further analysis & 160 \\
leftward axis & 435 \\
possible left anterior fascicular block & 138 \\
rightward axis & 164 \\
probable left atrial enlargement & 224 \\
low qrs voltages in precordial leads & 540 \\
st junctional depression is nonspecific & 149 \\
possible inferior infarct - age undetermined & 425 \\
lateral t wave changes are nonspecific & 328 \\
short pr interval & 167 \\
inferior t wave changes are nonspecific & 312 \\
left ventricular hypertrophy & 428 \\
lvh with secondary repolarization abnormality & 285 \\
left axis deviation & 1067 \\
poor r wave progression - probable normal variant & 538 \\
indeterminate axis & 108 \\
possible anterior infarct - age undetermined & 511 \\
anterior t wave changes are nonspecific & 182 \\
possible left atrial abnormality & 271 \\
inferior/lateral st-t changes are nonspecific & 240 \\
prolonged qt interval & 618 \\
possible anteroseptal infarct - age undetermined & 254 \\
septal t wave changes are nonspecific & 134 \\
right bundle branch block & 517 \\
lateral st-t changes are nonspecific & 289 \\
anteroseptal infarct - age undetermined & 129 \\
left anterior fascicular block & 202 \\
extensive st-t changes are nonspecific & 111 \\
inferior infarct - age undetermined & 550 \\
rsr'(v1) - probable normal variant & 199 \\
left bundle branch block & 354 \\
low qrs voltages in limb leads & 395 \\
extensive st-t changes may be due to myocardial ischemia & 143 \\
possible left ventricular hypertrophy & 150 \\
abnormal r-wave progression, early transition & 102 \\
inferior infarct, old & 123 \\
ventricular premature complex & 119 \\
possible septal infarct - age undetermined & 188 \\
right axis deviation & 141 \\
lateral st-t changes may be due to myocardial ischemia & 227 \\
inferior/lateral st-t changes may be due to myocardial ischemia & 167 \\
iv conduction defect & 376 \\
generalized low qrs voltages & 161 \\
qrs changes v3/v4 may be due to lvh but cannot rule out anterior infarct & 103 \\
lateral t wave changes may be due to myocardial ischemia & 106 \\
rbbb with left anterior fascicular block & 314 \\
extensive st-t changes may be due to hypertrophy and/or ischemia & 135 \\
normal ecg & 753 \\
normal ecg except for rate & 334 \\
abnormal ecg & 4761 \\
borderline ecg & 2074 \\
inferior/lateral st-t changes may be due to hypertrophy and/or ischemia & 116 \\
lateral st-t changes may be due to hypertrophy and/or ischemia & 112 \\
\end{longtable}
\endgroup

Figure~\ref{fig:cxr_report} presents an example report from the MIMIC-IV-CXR dataset, used in constructing CXR-QA. The report contains some unclear and unrelated content to the CXR image description. By applying the previously described CXR prompts, we transformed the report into a more organized and focused description centered on CXR.

\begin{center}
\tiny
\captionsetup{justification=centering} 
\captionof{figure}{An CXR report example from MIMIX-IV-CXR-report dataset.} 
\label{fig:cxr_report}
\fbox{
\begin{minipage}{0.9\textwidth}
\begin{center} 
\textbf{FINAL REPORT}\\
\end{center}
\vspace{0.5em}
\textbf{EXAMINATION:} CHEST (PA AND LAT)\\
\vspace{0.5em}
\textbf{INDICATION:} \_\_\_F with new onset ascites \slash\slash\ eval for infection\\
\vspace{0.5em}
\textbf{TECHNIQUE:} Chest PA and lateral\\
\vspace{0.5em}
\textbf{COMPARISON:} None.\\
\vspace{0.5em}
\textbf{FINDINGS:}
There is no focal consolidation, pleural effusion or pneumothorax. Bilateral nodular opacities that most likely represent nipple shadows. The cardiomediastinal silhouette is normal. Clips project over the left lung, potentially within the breast. The imaged upper abdomen is unremarkable. Chronic deformity of the posterior left sixth and seventh ribs are noted.\\
\vspace{0.5em}
\textbf{IMPRESSION:}
No acute cardiopulmonary process.\\
\end{minipage}
}
\end{center}

Disease-level labels are derived from the MIMIC-IV-ECG-EXT-ICD~\cite{MIMIC-IV-ECG-Ext-ICD} dataset, with these labels stored as ICD-10 codes. Each sample may correspond to multiple disease categories. We filtered out diseases for which evidence could not be found in ECG, CXR, or LAB data. Ultimately, we identified seven main categories: Coronary Artery Disease, Acute Renal Failure, Hypertension, Atrial Fibrillation, Pneumonia, Diabetes Mellitus, and Sepsis, along with some subclasses within these categories. Details are provided in Table~\ref{tab:ICD-10}.

\begingroup
\scriptsize
\begin{longtable}{|l|l|r|}
\caption{ICD-10 Disease Statistics with Corresponding Counts.} \\
\hline
\label{tab:ICD-10}
\textbf{Disease Category} & \textbf{ICD-10 Code} & \textbf{Count} \\
\hline
\endfirsthead
\hline
\textbf{Disease Category} & \textbf{ICD-10 Code} & \textbf{Count} \\
\hline
\endhead
\hline
\endfoot
\hline
\endlastfoot
\multicolumn{3}{|l|}{\textbf{Coronary Artery Disease}} \\
Coronary Artery Disease & I2510 & 2680 \\
Chronic ischemic heart disease, unspecified & I252  & 936  \\
Atherosclerotic heart disease of native coronary artery & I259  & 190  \\
Other forms of chronic ischemic heart disease & I253  & 8    \\
Ischemic cardiomyopathy & I255  & 79   \\
\hline
\multicolumn{3}{|l|}{\textbf{Acute Renal Failure}} \\
Acute kidney failure, unspecified & N179  & 2379 \\
Acute kidney failure with tubular necrosis & N170  & 689  \\
Acute kidney failure with other specified morphologic lesions & N178  & 12   \\
Acute kidney failure with acute cortical necrosis & N171  & 1    \\
\hline
\multicolumn{3}{|l|}{\textbf{Hypertension}} \\
Essential (primary) hypertension & I10   & 4155 \\
Hypertensive heart and chronic kidney disease, unspecified & I129  & 1536 \\
Hypertensive heart disease with heart failure & I120  & 515  \\
Hypertensive heart and chronic kidney disease with heart failure & I130  & 119  \\
Hypertensive heart disease without heart failure & I110  & 77   \\
Hypertensive heart and chronic kidney disease with heart failure and stage 5 CKD or ESRD & I132  & 37   \\
Hypertensive heart disease, unspecified & I119  & 8    \\
Other secondary hypertension & I159  & 1    \\
Renovascular hypertension & I150  & 7    \\
Other specified secondary hypertension & I158  & 1    \\
\hline
\multicolumn{3}{|l|}{\textbf{Atrial Fibrillation}} \\
Persistent atrial fibrillation & I4891 & 2623 \\
Permanent atrial fibrillation & I4892 & 317  \\
Paroxysmal atrial fibrillation & I480  & 237  \\
Other specified atrial fibrillation & I482  & 104  \\
Atrial flutter, unspecified & I481  & 15   \\
Typical atrial flutter & I483  & 1    \\
Atypical atrial flutter & I484  & 1    \\
\hline
\multicolumn{3}{|l|}{\textbf{Pneumonia}} \\
Pneumonia, unspecified organism & J189  & 1442 \\
Pneumonia due to other specified bacteria & J181  & 25   \\
Pneumonia due to Haemophilus influenzae & J188  & 18   \\
Pneumonia due to Klebsiella pneumoniae & J180  & 5    \\
\hline
\multicolumn{3}{|l|}{\textbf{Diabetes Mellitus}} \\
Type 2 diabetes mellitus without complications & E119   & 1752 \\
Type 2 diabetes mellitus with unspecified complications & E1129  & 217  \\
Type 2 diabetes mellitus with hyperglycemia & E11319 & 296  \\
Type 2 diabetes mellitus with renal complications & E1140  & 348  \\
Type 2 diabetes mellitus with circulatory complications & E1165  & 360  \\
Type 2 diabetes mellitus with other specified complications & E118   & 29   \\
Type 2 diabetes mellitus with unspecified complications & E139   & 29   \\
Type 1 diabetes mellitus without complications & E109   & 27   \\
Type 1 diabetes mellitus with unspecified complications & E138   & 2    \\
Type 1 diabetes mellitus with other specified complications & E108   & 3    \\
\hline
\multicolumn{3}{|l|}{\textbf{Sepsis}} \\
Sepsis, unspecified organism & A419  & 1088 \\
Severe sepsis without septic shock & R6520 & 939  \\
Severe sepsis with septic shock & R6521 & 873  \\
Sepsis due to Streptococcus pneumoniae & A403  & 26   \\
Sepsis due to Escherichia coli & A412  & 2    \\
Sepsis due to other Gram-negative organisms & A409  & 69   \\
Sepsis due to anaerobes & A414  & 28   \\
Sepsis due to other specified bacteria & A411  & 26   \\
Sepsis due to Staphylococcus aureus & A401  & 3    \\
Sepsis due to other specified Staphylococcus & A408  & 4    \\
Sepsis due to other specified bacteria & A413  & 1    \\
Sepsis due to unspecified Staphylococcus & A400  & 2    \\
\end{longtable}
\endgroup

\section{Examples of QA Pairs Corresponding to Each Modality}
\label{sec:QA_pairs}

This section provides examples of question-answer pairs corresponding to each modality (e.g., ECG, CXR, LAB) in the MedTVT-QA dataset.

\begin{tcolorbox}[colframe=gray, colback=gray!20, boxrule=0.5mm, title={ECG-QA Example}]
\scriptsize
\textbf{\textit{Question}:} What information can be derived from the ECG <ecg>?  \\
\textbf{\textit{Answer}:} ECG Analysis: Upon evaluating the electrocardiogram (ECG), the following observations can be made: \\
1.Sinus Rhythm: The heart is under the control of the sinoatrial (SA) node, which is the natural pacemaker. This finding suggests that the electrical activity of the heart is originating from the SA node and progressing through the normal conduction pathway, leading to a regular rhythm. This is a foundational indicator of organized cardiac activity.\\
2. Left Bundle Branch Block (LBBB): There is evidence of delayed or disrupted conduction through the left bundle of the heart's electrical conduction system. This results in a characteristic pattern on the ECG, with a widened QRS complex (>120 ms) and specific morphologies in leads, such as a broad, notched appearance in V5-V6 and an absence of normal septal depolarization. LBBB may occur in conditions like ischemic heart disease, cardiomyopathy, or hypertension-induced structural changes, though it can also occasionally be found in asymptomatic individuals.\\
Summary: The heart demonstrates a sinus rhythm, indicating normal initiation of electrical activity. However, the presence of a left bundle branch block signifies an abnormality in electrical conduction, often associated with underlying cardiac pathology. As a whole, this ECG is classified as abnormal due to the conduction disturbance, warranting further investigation to assess structural or functional cardiac issues.\\
\end{tcolorbox}

\begin{tcolorbox}[colframe=gray, colback=gray!20, boxrule=0.5mm, title={CXR-QA Example}]
\scriptsize
\textbf{\textit{Question}:} Would you mind explaining the findings on my chest X-ray <cxr>?\\
\textbf{\textit{Answer}:} 
Based on the evaluation of your chest X-ray image, the lungs appear overall unchanged from prior assessments, indicating no new substantial developments. There is no evidence of acute cardiopulmonary disease, which means there are no significant new concerns regarding your heart or lungs. Specifically, there are no signs of pneumonia, which would typically appear as clouded or consolidated areas in the lung fields. Similarly, there is no vascular congestion; this would generally be associated with heart-related problems affecting the lung's blood vessels and is absent in this image. Additionally, there is no pleural effusion, which would be fluid accumulation in the space between the lungs and chest wall.\\
However, there is a notable finding of a dense streak of atelectasis at the left base of the lung, which is a condition where a part of the lung collapses or does not inflate properly, resulting in that portion not being fully functional. This streak appears denser than usual tissue, indicating that a segment of the lung at the left base may not be fully expanded or aerated. Atelectasis can result from various causes, such as mucus blocking an airway, external pressure on the lung, or post-surgical changes.\\
Despite this abnormality, the absence of acute cardiopulmonary disease indicators is reassuring. The streak of atelectasis should be monitored, and clinical correlation with your symptoms is important. Continuous observation or additional imaging might be recommended to ensure there are no underlying issues contributing to this finding. Always consult with your healthcare provider to discuss the implications of these findings based on your clinical history and symptoms.
\end{tcolorbox}

\begin{tcolorbox}[colframe=gray, colback=gray!20, boxrule=0.5mm, title={LAB-QA Example}]
\scriptsize
\textbf{\textit{Question}:} Would you be able to undertake a comprehensive examination of this patient's blood panel <lab>, including the reference ranges and clinical implications associated with all abnormal values?\\
\textbf{\textit{Answer}:} Comprehensive Blood Panel Analysis:This blood test reflects multiple systemic functions. I will analyze the provided values across seven categories, indicating normal findings and clinical implications for abnormal or missing data.\\
Routine Blood Indicators: Hematocrit, hemoglobin, red blood cells, MCV, and MCHC fall within normal ranges, suggesting typical oxygen-carrying capacity and erythrocyte health. RDW is borderline high, which could indicate subtle variability in red cell size, potentially pointing toward early nutritional deficiencies or mild anemia. Platelet count and white blood cells are normal, showing adequate clotting potential and immune activity. All differential leukocyte percentages are within normal ranges, reflecting balanced immunity.\\
Electrolyte and Metabolic Indicators: Potassium, sodium, magnesium, bicarbonate, and phosphate are within normal ranges, indicating balanced electrolytes and acid-base homeostasis. However, chloride is mildly low (97.0), potentially reflecting minor electrolyte imbalance or compensatory metabolic alkalosis. Glucose is significantly elevated (185.0), possibly suggesting poorly controlled diabetes mellitus or acute hyperglycemia. Lactate is elevated (4.6), which could suggest tissue hypoxia or metabolic distress. Calcium is borderline low (8.5), potentially indicating subtle hypocalcemia warranting further evaluation.\\
Renal Function Indicators: Creatinine is within the high-normal range, which might raise concerns for early renal insufficiency depending on age/gender. Urea nitrogen is elevated, possibly indicating dehydration, increased protein metabolism, or mild renal dysfunction.\\
Liver Function Indicators: ALT, AST, bilirubin, and alkaline phosphatase values are missing and assumed normal, precluding further analysis.\\
Acid-Base Balance and Gas Exchange: With missing pH and gas exchange values, bicarbonate is normal but anion gap is elevated (22.0), potentially indicating metabolic acidosis from lactate accumulation or unmeasured ions.\\
Coagulation Function Indicators: INR, PT, and PTT are missing and presumed normal, suggesting no overt coagulation abnormalities.\\
Other Indicators: Data gaps in albumin, base excess, and other parameters limit full metabolic insight.\\
Summary: Elevated glucose and lactate suggest metabolic dysregulation. Elevated urea nitrogen and borderline creatinine warrant renal assessment. Mild hypocalcemia and low chloride need further exploration. Elevated anion gap hints at underlying acidosis. Follow-up tests are recommended for confirmation and diagnosis refinement.
\end{tcolorbox}

\section{Performance of Various MLLMs in Handling Medical Reasoning and Diagnosis}
\label{sec:performance}

\begin{figure}[t]
\centering
\includegraphics[width=1\linewidth]{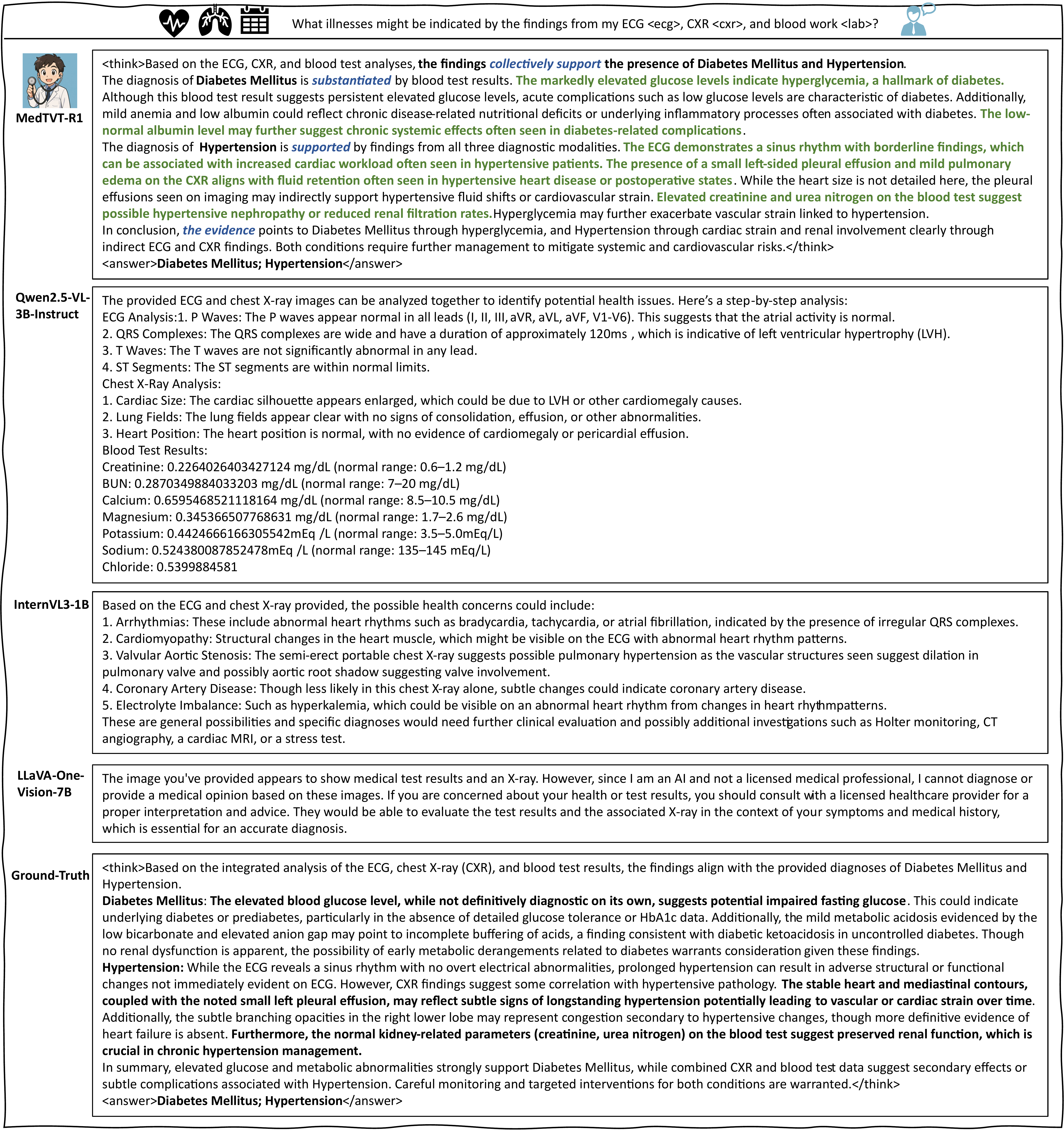}
\vspace{-1em}
  \caption{Performance of various MLLMs in handling medical reasoning and diagnosis. MedTVT-R1 approaches from the perspective of illness, tracing and identifying evidence of related diseases using the provided ECG, CXR, and LAB information. Qwen2.5-VL-3B-Instruct can analyze different modalities but lacks accuracy and cannot summarize diseases. InternVL3-1B combines modality information to determine disease types, though its accuracy is lacking. LLaVA-One-Vision-7B refuses to answer questions.}
\label{fig:supp_QA}
\end{figure}

Here, we present an example of medical analysis and diagnosis among various MLLMs and MedTVT-R1, as shown in Figure~\ref{fig:supp_QA}.

\section{Limitation}
\label{appendix:limitation}
Although our proposed MedTVT-R1 successfully integrates CXR, ECG, and LAB data for joint multi-disease diagnosis, there are still some limitations. First, precise disease diagnosis often requires a larger volume of multimodal data collected from the same patient within a similar timeframe. However, in the short term, it is challenging to gather such large-scale data, which limits the model’s generalization ability and diagnostic accuracy. Second, while our work incorporates three modalities—CXR, ECG, and LAB—more accurate disease analysis and diagnosis may rely on additional modalities, such as patient medical history, genomic data, or other biomarkers. Unfortunately, the current open-source datasets lack richer multimodal information, making such extensions difficult to achieve. Future research could aim to address these challenges to further enhance the model’s diagnostic capabilities and practical applicability.

\end{document}